\documentclass[11pt,a4paper]{report}
 \pdfoutput=1
\usepackage[utf8]{inputenc}
\usepackage[T1]{fontenc}
\usepackage[english]{babel}
\usepackage{amsmath,subfig}
\usepackage{ae}
\usepackage{icomma}
\usepackage{rotating}
\usepackage[table]{xcolor}
\usepackage{booktabs}
\usepackage{units}
\usepackage{color}
\usepackage{graphicx}
\usepackage{hvfloat}
\usepackage{epstopdf}
\usepackage{amsmath}
\usepackage{lipsum}
\usepackage{cases}
\usepackage{stfloats}
\usepackage{graphics}
\usepackage{verbatim}
\usepackage{amssymb}
\usepackage{float}
\usepackage{dsfont}
\usepackage{ifpdf}
\usepackage{cite}
\usepackage{hyperref}
\usepackage{caption}
\usepackage{multirow}
\usepackage{array}
\usepackage{geometry}
\usepackage{fancyhdr}
\usepackage{nomencl}

\usepackage[square, numbers, comma, sort&compress]{natbib}
\usepackage{verbatim}
\usepackage{acronym}
\usepackage{fncychap}
\usepackage{lettrine}
\usepackage{eso-pic}
\usepackage{mathtools,soul}
\usepackage{acronym}

\newcommand{\setdefaulthdr}{%
\fancyhead[L]{\slshape \rightmark}%
\fancyhead[R]{\slshape \leftmark}%
\fancyfoot[C]{\thepage}%
}
\newcommand{\setspecialhdr}{%
\fancyhead[L]{ }%
\fancyhead[R]{\slshape \leftmark}%
\fancyfoot[C]{\thepage}%
}

\newcommand{\mail}[1]{\href{mailto:#1}{\nolinkurl{#1}}}
\newcommand{\backgroundpic}[3]{%
	\put(#1,#2){
		 \parbox[b][\paperheight]{\paperwidth}{%
			\centering
			 \includegraphics[width=\paperwidth,height=\paperheight,keepaspectratio]{#3}
			\vfill
}}}

\hypersetup{
		pdftitle={Master's Thesis: },%
		pdfauthor={Arif Onder Isikman},%
    colorlinks=true,%
    citecolor=black,%
    filecolor=black,%
    linkcolor=black,%
    urlcolor=black
}

\makeatletter
\ChNumVar{} 
\ChTitleVar{\Huge\bfseries\centering} 
\ChRuleWidth{4pt} 
\ChNameUpperCase 
\renewcommand{\DOCH}{
\centering
{\CNoV {\fontsize{60pt}{20pt}\selectfont\thechapter} }
\vskip 40\p@}
\renewcommand{\DOTI}[1]{%
\CTV\FmTi{#1}\par\nobreak
\vskip 40\p@}
\renewcommand{\DOTIS}[1]{%
\CTV\FmTi{#1}\par\nobreak
\vskip 40\p@}
\makeatother

{\begin{center} \bfseries \abstractname \end{center}}%
{\vspace{2\baselineskip}}%

\setdefaulthdr

 \makenomenclature

\begin{document}

\begin{titlepage}

\AddToShipoutPicture{\backgroundpic{-4}{56.7}{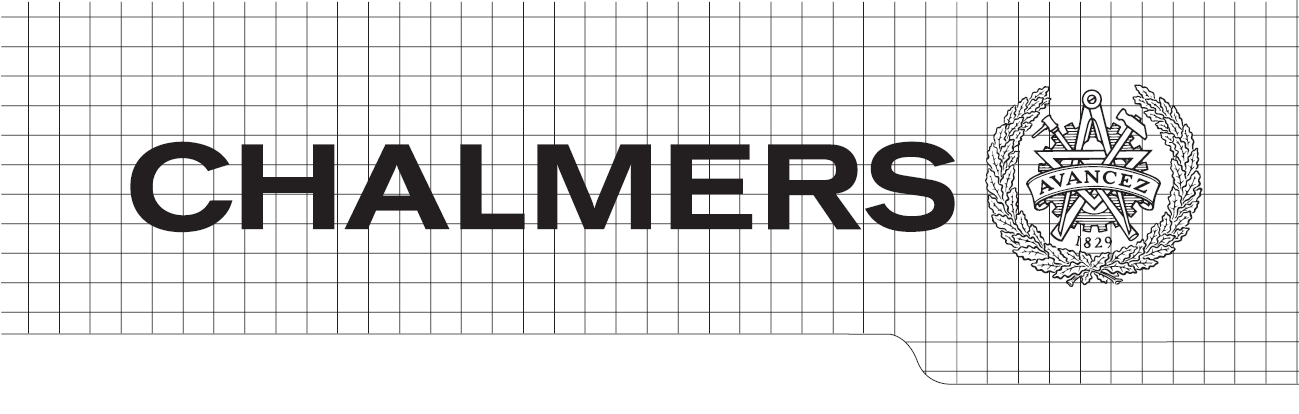}}
\vspace*{3cm}
{\hspace{-3cm}{\includegraphics[width=20cm]{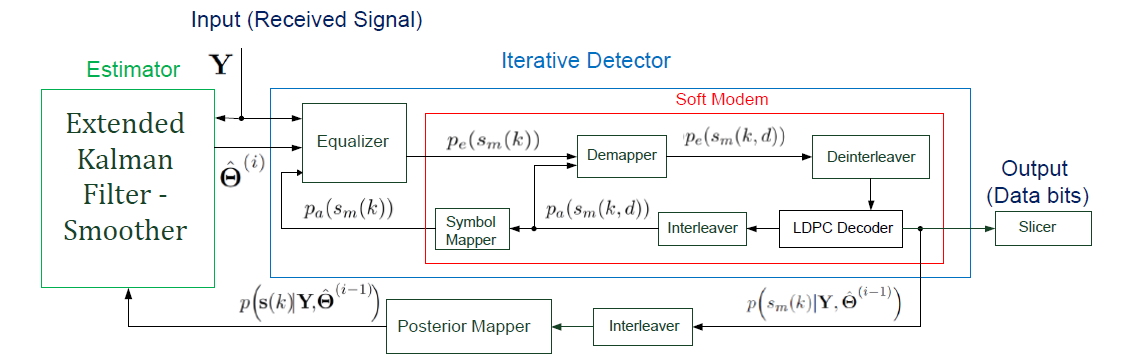}}}
\mbox{}
\vfill
\addtolength{\voffset}{2cm}
\begin{flushleft}
	{\noindent {\Huge Phase Noise Estimation for Uncoded/Coded SISO and MIMO Systems} \\[0.5cm]
	\emph{\Large Master's Thesis in the program Communication Engineering} \\[.8cm]
	
	{\huge ARIF ONDER ISIKMAN}\\[.8cm]
	
	{\Large Department of Signal \& Systems \\
	\textsc{Chalmers University of Technology} \\
	Gothenburg, Sweden 2012 \\
	Master's Thesis EX061/2012
	}
	}
\end{flushleft}

\end{titlepage}
\ClearShipoutPicture

\pagestyle{empty}
\newpage
The Author grants to Chalmers University of Technology the non-exclusive right to publish the Work electronically and in a non-commercial purpose make it accessible on the Internet. The Author warrants that he/she is the author to the Work, and warrants that the Work does not contain text, pictures or other material that violates copyright law.

The Author shall, when transferring the rights of the Work to a third party (for example a publisher or a company), acknowledge the third party about this agreement. If the Author has signed a copyright agreement with a third party regarding the Work, the Author warrants hereby that he/she has obtained any necessary permission from this third party to let Chalmers University of Technology store the Work electronically and make it accessible on the Internet.\\
\newline
\newline

Phase Noise Estimation for Uncoded/Coded SISO and MIMO Systems\\
ARIF ONDER ISIKMAN
\\
\\
\copyright \;ARIF ONDER ISIKMAN,\,2012
\\
\\
Examiner:\;  Assistant Professor Alexandre Graell i Amat
\\
\\
\\
Technical report no EX061/2012 \\
Department of Signals \& Systems\\
Chalmers University of Technology\\
SE-41296 Gothenburg\\
Sweden\\
Telephone +46 (0) 31-704 195422
\newline
\newline
\newline
\newline
\newline
\newline
\newline

\newpage
\clearpage
\mbox{}
\thispagestyle{empty}
\section*{Abstract}
 New generation cellular networks have been forced to support high data rate communications. The demand for high bandwidth data services has rapidly increased with the advent of bandwidth hungry applications.  To fulfill the bandwidth requirement, high throughput backhaul
links are required. Microwave radio links operating at high frequency bands  are used to fully exploit the available spectrum. Generating high carrier frequency becomes problematic due to the hardware limitations. Non-ideal  oscillators both at the transmitter and the receiver introduces time varying phase noise which interacts with the transmitted data in a non-linear fashion. Phase noise becomes a detrimental problem in digital communication systems and needs to be estimated and compensated. In this thesis receiver algorithms are derived and evaluated to mitigate the effects of the phase noise in digital communication systems.

The thesis is organized as follows: In Chapter \ref{chap:SISO-model} phase noise estimation in single-input single-output (SISO) systems is investigated. First, a hard decision directed extended Kalman filter (EKF) is derived and applied to track time varying phase noise for an uncoded system. Next, the problem of phase noise estimation for coded SISO system is investigated. An iterative receiver algorithm performing code-aided turbo synchronization is derived using the expectation maximization (EM) framework. Two soft-decision directed estimators in the literature based on Kalman filtering, the Kalman filter and smoother with maximum likelihood average (KS-MLA) and the extended Kalman filter and smoother (EKS),  are evaluated.  Low density parity check (LDPC) codes are proposed to calculate marginal a posteriori probabilities and to construct soft decision symbols. Error rate performance of both estimators, the KS-MLA and the EKS, are determined and compared through simulations. Simulations indicate that comparison on the performance of the existing estimators heavily depends on the system parameters such as block length and modulation order which are not taken into consideration in the literature.

In Chapter \ref{chap:MIMO-model} the thesis focuses on phase noise estimation in multi-input multi-output (MIMO) systems. MIMO technology is commonly used in microwave radio links to improve spectrum efficiency. First, an uncoded MIMO system is taken under consideration. A low complexity hard decision directed EKF is derived and evaluated.
A new MIMO receiver algorithm that iterates between the estimator and the detector, based on the EM framework for joint estimation and detection in coded MIMO systems in the presence of time varying phase noise is proposed. A low complexity soft decision directed extended Kalman filter and smoother (EKFS) that tracks the phase noise parameters over a frame is proposed in order to carry out the maximization step.
The proposed EKFS based approach is
combined with an iterative detector that utilizes bit interleaved coded modulation and employs LDPC codes to calculate the marginal a posteriori probabilities of the transmitted symbols, i.e., soft decisions.
Numerical investigations show that for a wide range of phase noise variances the estimation accuracy of the
proposed algorithm improves at every iteration. Finally, simulation results confirm that the error rate performance of the proposed EM-based approach is close to the scenario of perfect knowledge of
phase noise at low-to-medium signal-to-noise ratios.

\newpage
\clearpage

\mbox{}
\newpage
\clearpage
\thispagestyle{empty}
\section*{Acknowledgements}
I would first like to thank my supervisors Hani Mehrpouyan and Alexandre Graell i Amat for their support and guidance during the whole thesis process. I highly appreciate their attitude towards me. They have taught  me how an academic should approach the problems in the field of research. They have also trusted me and given me a lot of freedom. They respect my independent and somehow arrogant way of performing research. I would also like to thank my friends in Communication Engineering Masters Programme for the discussions and shared ideas. I want to also thank to the employees at Ericsson for showing me how things are done in industry in a short time.

Thanks also to my family and friends in Turkey for supporting me through all these years of studying. I have to thank my Swedish family, Jan and Ann-Charlotte Fonselius, for creating such a nice environment for me at the house which I share with my beloved friend Kiryl Kustanovich.

Last but not least, I want to thank Olric. I would not be able to finish this thesis without him.

\hfill Arif Onder Isikman, G\"{o}teborg \today
\newpage
\clearpage
\mbox{}

\newpage
\pagenumbering{roman}
\setcounter{page}{1}
\pagestyle{fancy}
\setspecialhdr
\tableofcontents
\listoffigures

\newpage
\setdefaulthdr
\pagenumbering{arabic}	
\setcounter{page}{1}

\chapter*{Acronym}

\begin{acronym}
\acro{APP}{a posteriori probability}
\acro{AWGN}{additive white gaussian noise}
\acro{BER}{bit error rate}
\acro{BICM}{bit interleaved coded modulation}
\acro{BS}{base station}
\acro{BSC}{base station controller}
\acro{DA}{data aided}
\acro{EKF}{extended Kalman filter}
\acro{EKS}{extended Kalman filter-smoother (SISO)}
\acro{EKFS}{extended Kalman filter-smoother (MIMO)}
\acro{EM}{expectation maximization}
\acro{FER}{frame error rate}
\acro{KS}{Kalman filter-smoother (SISO)}
\acro{KS-MLA}{Kalman filter-smoother with maximum likelihood average}
\acro{LDPC}{low-density parity-check}
\acro{LOS}{line-of-sight}
\acro{LLF}{log likelihood function}
\acro{LLR}{log likelihood ratio}
\acro{LS}{least square}
\acro{MAP}{maximum a posteriori}
\acro{MIMO}{multi-input multi-output}
\acro{ML}{maximum likelihood}
\acro{MMSE}{minimum mean square error}
\acro{MSC}{master switching center}
\acro{MSE}{mean square error}
\acro{QAM}{quadrature amplitude modulation}
\acro{SDMA}{space division multiple access}
\acro{SISO}{single-input single-output}
\acro{SNR}{signal to noise ratio}
\acro{VCO}{voltage controlled oscillator}
\acro{WLS}{weighted least square}

 \end{acronym}

\chapter{Introduction}\label{chap:intro}
 In recent years the demand for high bandwidth data services has increased with the evolution of the third generation (3G) and fourth generation (4G) cellular networks \cite{Boch}. Rapid escalation in the use of bandwidth hungry devices also increases the throughput requirements of the base station (BS), base station controller (BSC) and master switching center (MSC), which are the fundamental components of a cellular network.  The user connects to the network through the BS. Each BS is connected to a BSC via a wired or a wireless link. The BSC routes the data from the BS to the MSC and controls the functionality of the BS. The MSC holds all the network information and controls all calls and  data management functionalities. In other words, the MSC is the brain of any cellular network. The portion of a wireless mobile network from the BS to the MSC is called as \emph{backhaul network}.

 The backhaul links serves the medium to carry traffic from the BS to the MSC via the BSC. The point-to-point microwave radio links are commonly used in backhaul networks. They are cost efficient and can be deployed rapidly. Microwave radio transmission is operated at certain frequency bands. Lower bands such as 7, 18, 23 and 35GHz have better radio propagation characteristics. On the other hand, these frequency bands fail to provide sufficient bandwidth since the spectrum is mostly allocated. With the release of the E-Band, 10GHz of bandwith  in the spectrum at 70GHz (71-76GHz) and 80GHz (81-86GHz) have been made available for point-to-point microwave links. To meet high data rate requirements point-to-point microwave systems are equipped with multiple transmit and multiple receive antennas. \emph{Line-of-sight} (LOS) \emph{multi-input multi-output} (MIMO) systems are effectively used for backhaul networking \cite{article_MIMO_LoS}.

  Local oscillators are utilized to carry the baseband signal to the operating band. Due to the hardware limitations, every oscillator suffers from an instability of its phase, resulting in phase noise \cite{Demir}. Phase noise can dramatically limit the performance of a wireless communication system if left unaddressed
\cite{Meyr1997}. Phase noise interacts with the transmitted symbols both at the transmitter and the receiver side in a non-linear manner and significantly distorts the received signal. Digital signal processing algorithms need to be employed to achieve synchronous transmission in the presence of phase noise.  Several algorithms are proposed for \emph{single-input single-output }(SISO) systems to mitigate the effect of time varying phase noise \cite{Colavolpe,Godtmann,Simon,Shehata,Bhatti}. In the case of LOS-MIMO systems, each transmit and receive antenna is equipped with a different oscillator since the antennas are placed far apart. Similarly, in the case of multi-user MIMO systems or \emph{space division multiple access} (SDMA) systems independent oscillators are used by different users  to transmit their data to common receiver \cite{article_PHASE_N_MIMO_OFDM_VIII}. As a result, a single oscillator cannot be employed and phase noise compensation algorithms proposed for SISO systems are not directly applicable to MIMO systems.


\section{Background}

Achieving channel capacity was seen far from reality until two decades ago. The introduction of turbo codes \cite{Berrou} and the rediscovery of \emph{low-density-parity-check (LDPC)} codes \cite{Gallager} has demonstrated the power of the iterative processing paradigm in improving the performance of communication systems and in operating close to the theoretical limits. Subsequently, the iterative coding structure has been applied to facilitate and improve many functions including synchronization. Parameter estimation can be performed jointly with data detection in an iterative fashion. It is well-known that the application of turbo codes and LPDC codes improves the data detection process at the receiver, which in turn can be applied to improve the performance of decision-directed estimators. The improved estimation and tracking accuracy allows for more accurate compensation of impairments such as time varying phase noise at the receiver which can also improve data detection. Thus, by jointly performing data detection and estimation, the performance of wireless communication systems can be significantly improved. This approach, known as ``turbo synchronization", was initially proposed in \cite{Lottici} and has since been formalized in \cite{Noels} with the use of the \emph{expectation-maximization (EM)} framework \cite{Moon}.

In \cite{Herzet1}, different frameworks for turbo synchronization based on the gradient method and the sum-product algorithms are studied. This work is extended to the problem of estimation of time varying phase noise for SISO systems in \cite{Shehata}. In \cite{Shehata}, based on the assumption of small phase noise values within each block and removing the data dependency from the observed signal, the tracking is carried out via a modified EM-based algorithm that applies a soft decision-directed linearized Kalman Smoother. In addition, to enhance phase noise tracking performance for very high phase noise variances, \cite{Shehata} proposes to employ a maximum-likelihood (ML) estimator in conjunction with a Kalman smoother, labeled as (KS-MLA).  A soft decision-directed extended Kalman filter-smoother (EKS) is also suggested to provide phase noise estimation. However, the performance of the KS-MLA degrades with increasing block length. More importantly, the linearization applied in \cite{Shehata} is not applicable to MIMO systems and the estimation performance of the proposed tracking algorithm is not investigated.

MIMO technology allows communication systems to more efficiently use the available spectrum \cite{Telatar99},\cite{Foschini98}. \emph{Bit-interleaved-coded-modulation (BICM)} is one of the popular schemes that enables communication systems to fully exploit the spectrum efficiency promised by MIMO technology \cite{Duman,Boutros}. However, the performance of MIMO systems degrades dramatically in the presence of synchronization errors. Code-aided synchronization based on the EM framework for joint channel estimation, frequency and time synchronization for a BICM-MIMO system is proposed in \cite{HenkSimoens}. However, in \cite{HenkSimoens}, the synchronization parameters are assumed to be constant and deterministic over the length of a block which is not a valid assumption for time varying phase noise.

 A Wiener filter approach that applies spatial correlation to improve phase noise estimation in MIMO systems is proposed in \cite{Hadaschik}. However, the proposed solution is only applicable to uncoded MIMO systems and the algorithm in \cite{Hadaschik} introduces significant overhead to phase noise estimation process since it requires frequent transmission of \emph{orthogonal} pilot symbols. The problem of joint data detection and phase noise estimation for coded MIMO systems over block fading channels is still unaddressed and will be the main focus of this thesis.

\section{Thesis Organization}

 In Chapter 2 the phase noise model is introduced and digital communication system for SISO systems over the additive white Gaussian noise (AWGN) channel affected by phase noise is presented.

 In Chapter 3 the performance of both uncoded and coded SISO systems affected by phase noise are investigated. The iterative code-aided EM-based approach used in \cite{Shehata} is modified and derived analytically. The EM-based algorithm is implemented and its components are explained in detail. Two estimators that are proposed in \cite{Shehata}, the KS-MLA and the EKS, are evaluated and their performances are compared against one another.

 In Chapter 4, the MIMO system model for both uncoded and the coded MIMO systems over Rician fading channels in the presence of phase noise is described in detail. An iterative joint phase noise estimation and data detection algorithm based on the EM framework is derived analytically. A low complexity \emph{extended Kalman filter-smoother} (EKFS) is proposed to estimate the time varying phase noise processes of each oscillator. BICM scheme is used to decrease the detection complexity. The performance of the proposed algorithm is investigated via computer simulations.

 In Chapter 5 conclusion and future research directions  are discussed.

\section{Thesis Contributions}
The primary contributions of this thesis are summarized as follows:

\begin{itemize}
\item The system model for both uncoded SISO and uncoded MIMO systems in the presence of phase noise are  outlined in detail and an extended Kalman filter with symbol-by-symbol feedback is proposed for each system. The error rate performance of the proposed estimators are investigated through numerical results.
\item The iterative code-aided EM-based algorithm proposed in \cite{Shehata} for coded SISO systems is modified and derived analytically. Moreover, the performances of two estimators, the KS-MLA and the EKS, are numerically compared with the help of computer simulations.
\item An EM-based receiver is proposed to perform iterative joint phase noise estimation and data detection for BICM-MIMO systems.
\item It is analytically demonstrated that a computationally efficient EKFS can be applied to carry out the maximization step of the EM algorithm.
\item A new low complexity soft decision-directed EKFS for tracking phase noise over the length of a frame is proposed and the filtering and smoothing equations are derived.
 \item Extensive simulations are carried out for different phase noise variances to show that the performance of a MIMO system employing the proposed receiver structure is very close to the ideal case of perfect knowledge of phase noise.  Simulation results demonstrate that error rate performances of a 2$\times$2 LOS-MIMO system using the proposed EM-based receiver is very close to that of the perfectly synchronized system for low-to-medium signal-to-noise ratios. It is also shown that the \emph{mean square-error }(MSE) of the phase noise estimates improves with every EM iteration.
\end{itemize}

\chapter{Digital Communications \\ with Phase Noise}\label{chap:theory-model}
\section{Phase Noise Modeling}
In wireless communication systems, the baseband signal is multiplied by a high frequency sine wave to operate at a certain frequency band, called carrier frequency. Local oscillators produce the carrier frequency waveforms. Phase-Locked Loop (PLL) \cite{Gardner} is the major phase recovery block in a communication system. PLL calculates the phase difference between the input and the output signal. The difference is then filtered by a low pass filter and applied to the voltage controlled oscillator (VCO). The controlled voltage on the VCO changes the oscillator frequency to minimize the phase difference of the input and the output signal. However, the output of the VCO circuit is a non-ideal sine wave due to some hardware limitations.  The power spectrum of the output signal is not strictly concentrated at the carrier frequency. The instantaneous output of a oscillator is given by \cite{eric04}
\begin{eqnarray}
V(t)=V_0(1+A(t))\exp\Big(j( 2 \pi f_c t +\theta(t))\Big)
\end{eqnarray}
where $f_c$ denotes the carrier frequency, $V_0$ denotes the amplitude, $A(t)$ is amplitude noise and $\theta(t)$ is phase noise. Demir \emph{et. al.} show in \cite{Demir} that amplitude noise decays over time, since the system stabilizes itself. The amplitude noise may thus be ignored and the normalized oscillator output signal can be written as
\begin{eqnarray}
V(t)=e^{(j 2 \pi f_c t)} e^{(j\theta(t))}.
\end{eqnarray}

The oscillator phase noise can be seen as a widening of the spectral peak of the oscillator. The frequency domain single-sideband phase noise power, $\mathcal{L}(f)$ is defined as the ratio of the noise power in a 1Hz sideband at an offset $f$ Hertz away from the carrier, $P_{SSB}$, to the total signal power, $P_{c}$.

Since we have no absolute time reference, the phase disturbances accumulate over time and can be represented by
\begin{eqnarray}
\theta(t)=\int^t_0\upsilon(s)ds \label{eqn:contPhaseNoise}
\end{eqnarray}
where $\upsilon(t)$ is a white Gaussian process with a constant power spectral density (PSD). Then, the phase noise process can be modeled as a  Wiener process and the oscillator power spectrum is a Lorentzian, given by \cite{Demir}
\begin{eqnarray}
\mathcal{L}(f)=\frac{1}{\pi f_{3dB}}\frac{1}{1+\Big(\frac{f} {f_{3dB}}\Big)^2}
\end{eqnarray}
where $f_{3dB}$ denotes the 3dB bandwidth. It is seen that the spectrum is characterized by a single parameter, $f_{3dB}$.
The phase noise process is sampled every $T_s$ seconds, sampling time interval. Then, the discrete time phase noise process is defined as,
\begin{eqnarray}
\theta(k)\triangleq\theta(kT_s),
\end{eqnarray}
and can be modeled as a random walk in accordance with \ref{eqn:contPhaseNoise}, i.e. discrete-time Wiener process \cite{Demir}
\begin{eqnarray}
\theta(k)=\theta(k-1)+\Delta(k). \label{eqn:thataDelta1}
\end{eqnarray}
In (\ref{eqn:thataDelta1}), the innovation term, $\Delta(k)$ is a discrete zero-mean Gaussian random variable with  variance $\sigma^2_\Delta$, denoted as $\mathcal{N}\left(0, \sigma^2_\Delta\right)$.
The phase noise innovation variance is given by \cite{Demir}
\begin{eqnarray}
\sigma^2_\Delta=4\pi f_{3dB} T_s.
\end{eqnarray}
Note that the discrete innovation process is also white,
\begin{eqnarray}
\mathbb{E}(\Delta(k)\Delta(l))=0, k\neq l.
\end{eqnarray}

\begin{figure}[htbp!]
\begin{center}
\includegraphics[width=10cm]{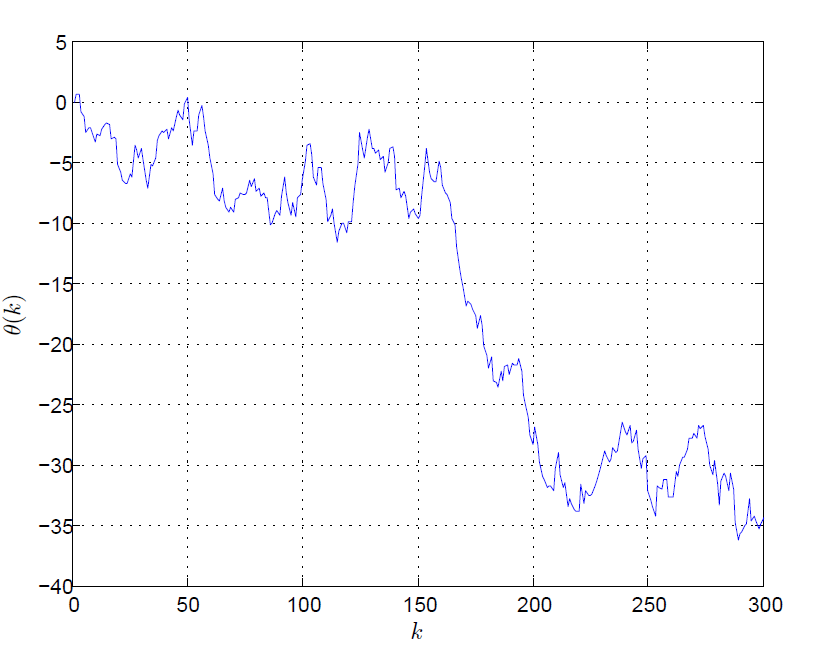}\center
\caption{Wiener phase noise for $\sigma^2_\Delta=1^\circ$.}\label{fig:figwienerPN}
\end{center}
\end{figure}

In Fig. \ref{fig:figwienerPN} a realization of the discrete time Wiener phase noise process is plotted.

\subsection{Digital Communications in the Presence of Phase Noise}
\begin{figure}[htbp!]
\begin{center}
\includegraphics[width=10cm]{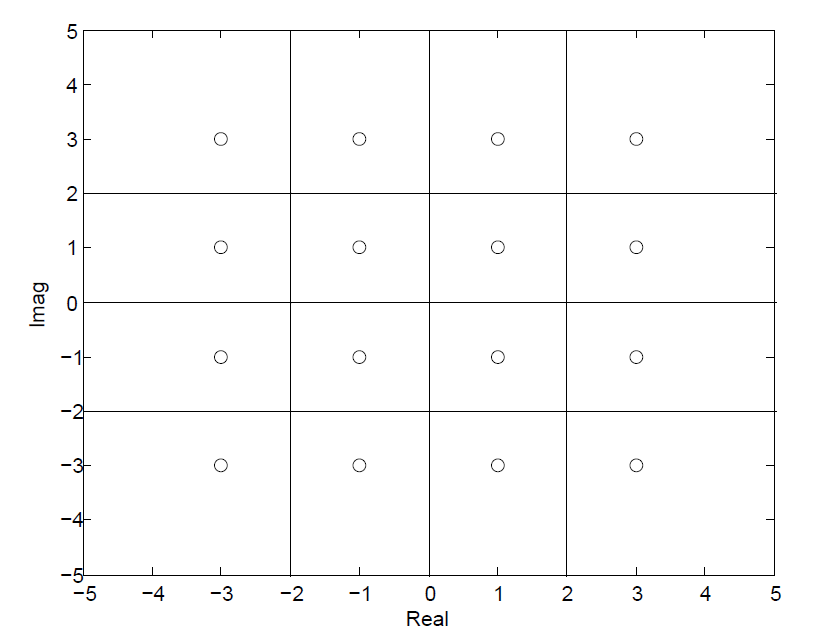}\center
\caption{16-QAM constellation.}\label{fig:fig16QAM}
\end{center}
\end{figure}

\begin{figure}[htbp!]
\begin{center}
\includegraphics[width=10cm]{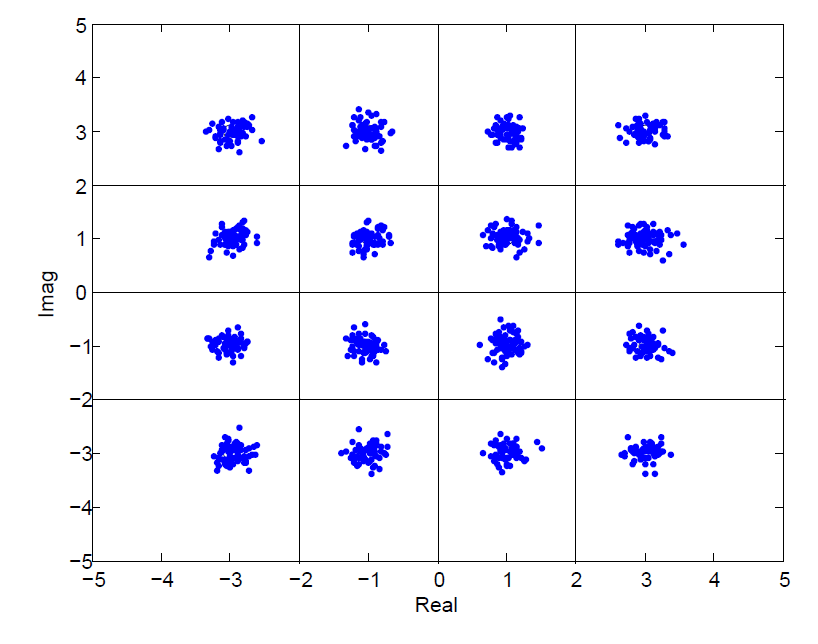}\center
\caption{16-QAM constellation affected by AWGN.}\label{fig:fig16QAMAWGN}
\end{center}
\end{figure}

At the transmitter, a group of data bits are modulated onto an $M$-point quadrature amplitude modulation ($M$-QAM) constellation $ \Omega$, displayed in Fig. \ref{fig:fig16QAM}. Symbols are then transmitted through an AWGN channel. In a communication system without the phase noise disturbances the received signal at time $k$ is given by
\begin{eqnarray}
 y(k)&=&s(k)+\tilde w(k)\label{eqn:recSign1}
\end{eqnarray}
where $y(k)$ is the received signal, $s(k)$ is the complex transmitted symbol, $\tilde w(k)$ is the zero-mean AWGN with variance $\sigma^2_w/2$ per dimension, i.e. $\tilde w(k)\sim\mathcal{N_C}\left(0, \sigma^2_w\right)$, as shown in Fig.  \ref{fig:fig16QAMAWGN}.

The received signal is also effected by time varying phase noise both at transmitter and receiver. Let $\theta^{[t]}(k)$ and $\theta^{[r]}(k)$ denote the discrete time phase noise sample at transmitter and receiver, respectively. The received signal at time $k$ is given by
\begin{eqnarray}
 y(k)&=&(s(k) e^{j\theta^{[t]}(k)}+\tilde w(k))e^{j\theta^{[r]}(k)}\\
     &=&s(k) e^{j(\theta^{[t]}(k)+\theta^{[r]}(k))}+ w(k)\label{eqn:recSign2}\\
      &=&s(k) e^{j\theta(k)}+w(k)\label{eqn:obsPN}
\end{eqnarray}
where $w(k)\triangleq \tilde w(k)e^{j\theta^{[r]}(k)}$ is the rotated noise sample, and $e^{j\theta(k)}$ is the total phase noise process. Note that rotation on the circular symmetric additive noise does not change the statistical properties, i.e., $ w(k)\sim\mathcal{N_C}\left(0, \sigma^2_w\right)$. The innovation of total phase noise process will have a zero-mean Gaussian distribution and its variance will be the sum of the variances, $ \Delta(k)\sim\mathcal{N}\left(0, \sigma^2_{\Delta^{[t]}}+\sigma^2_{\Delta^{[r]}}\right)$ where $\sigma^2_{\Delta^{[t]}}$ and $\sigma^2_{\Delta^{[r]}}$ denote the innovation variance of the phase noise process at the transmitter an at the receiver, respectively.

\begin{figure}[htbp!]
\begin{center}
\includegraphics[width=10cm]{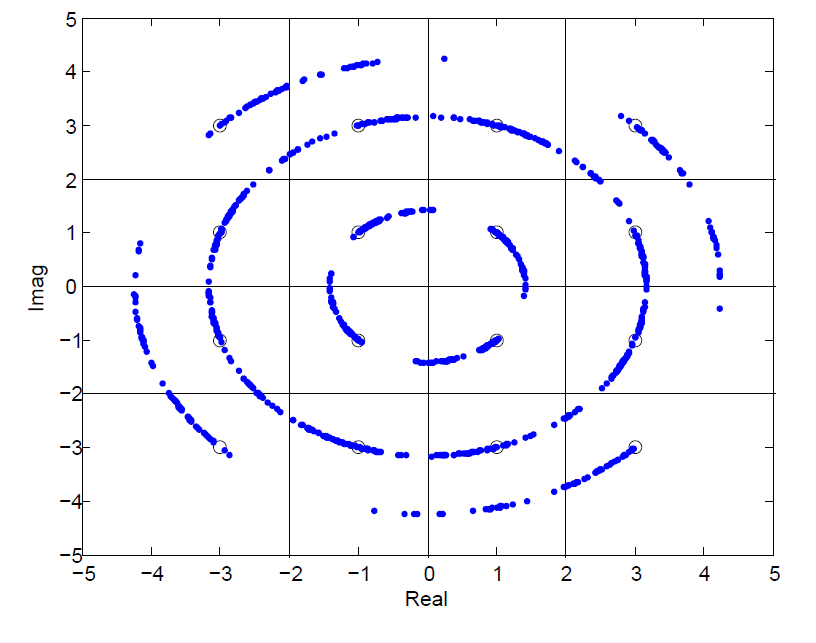}\center
\caption{16-QAM constellation rotated by phase noise.}\label{fig:fig16QAMPN}
\end{center}
\end{figure}

\begin{figure}[htbp!]
\begin{center}
\includegraphics[width=10cm]{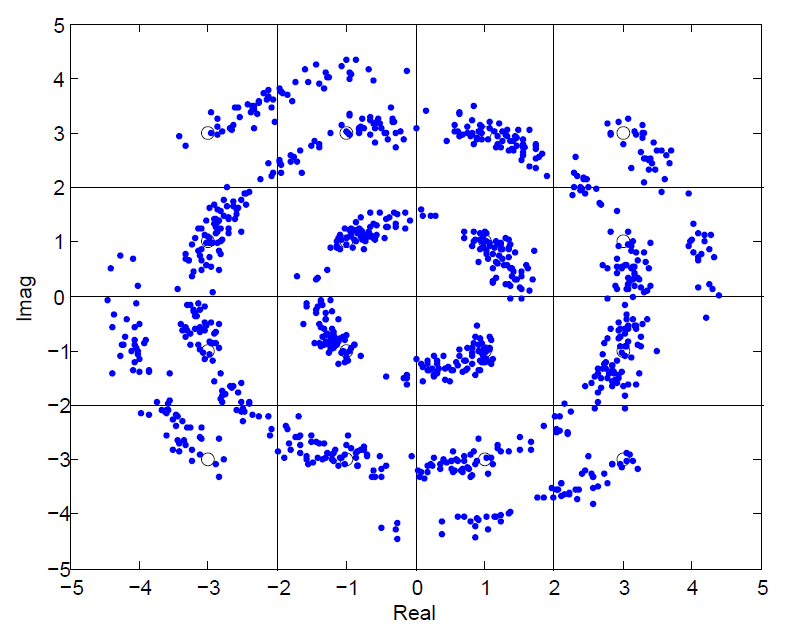}\center
\caption{Received signal affected by both phase noise and AWGN.}\label{fig:fig16QAMAWGNandPN}
\end{center}
\end{figure}

The total phase noise process rotates the signal constellation as displayed in Fig \ref{fig:fig16QAMPN}. The received signal which is affected by the AWGN and rotated by the phase noise is shown in Fig. \ref{fig:fig16QAMAWGNandPN}.

\chapter{Phase Noise Estimation \\ for SISO systems}\label{chap:SISO-model}
The main focus of this thesis will be on the phase noise estimation for coded MIMO systems. In this chapter, to better understand the effect of phase noise, the problem of  phase noise estimation for SISO systems over AWGN channel is investigated. It is assumed that perfect frame synchronization and phase recovery are performed at the beginning of each frame by transmitting sufficient number of pilot symbols. Therefore, the problem under consideration for SISO systems is simplified to the problem of phase noise estimation before investigating the problem for coded MIMO systems.
First, an uncoded SISO system is taken under consideration in Sec. \ref{sec:UncodedSISO}.   An extended Kalman filter (EKF) is suggested to track time varying phase noise and the set of equations for the EKF is derived. In Sec. \ref{sec:CodedSISO}, the problem of joint phase noise estimation and detection for a coded SISO system is discussed. An algorithm is analytically derived from the EM framework and is applied to iteratively solve the problem. The enhancement of the LDPC codes and the Kalman filters into the EM-based algorithm is explained in detail.

\section{Uncoded SISO}\label{sec:UncodedSISO}
In order to de-rotate the signal space and to achieve synchronous communication,  time varying phase noise process should be estimated. Since the parameter to be estimated is not deterministic, an estimator based on the Bayesian approach should be used \cite{SMKay}. In the Bayesian approach, prior knowledge about the random parameter is also taken into account. This approach is commonly used for the systems which can be represented with a dynamical model. Kalman filtering can be considered as a sequential minimum mean square error (MMSE) estimator which works according to the Bayesian framework.

    The received signal in (\ref{eqn:obsPN}) and the phase noise process are used to construct a state space signal model. The state and observation equations at time $k$ are given as
\begin{eqnarray}
 \textrm{State:}  &\theta(k)&=\theta(k-1)+\Delta(k),\\
  \textrm{Observation:}  &y(k)&=s(k) e^{j\theta(k)}+w(k),\label{eqn:obsUncoded}
\end{eqnarray}
where $y(k)$ is the received signal, $s(k)$ is the complex transmitted symbol belonging to the $M$-QAM constellation, $\theta(k)$ is the phase noise value, $w(k)$ is the additive complex Gaussian noise with variance $\sigma^2_w$ and $\Delta(k)$ is the phase noise innovation, which is assumed to be Gaussian distributed with variance $\sigma^2_\Delta$.
Since the observation equation is nonlinear, a hard decision-directed EKF is used instead \cite{SMKay}. The observation equation can be rewritten as
\begin{eqnarray}
y(k)&=& z(\theta(k))+w(k),
\end{eqnarray}
where the nonlinear function $z(.)$ is defined as
\begin{eqnarray}
z(\theta(k))\triangleq \hat s(k) e^{j\theta(k)}.\label{eqn:zFunc1}
\end{eqnarray}
In (\ref{eqn:zFunc1}), $\hat s(k)$ is the hard decision of the transmitted symbol at time instance $k$. Note that for uncoded SISO systems hard decision symbol, $\hat s(k)$,  is obtained by the demodulator at each time instance. Next, $\hat s(k)$ is input to the hard decision-directed EKF at each time instance. In other words, decision feedback is performed symbol-by-symbol. The EKF provides phase noise estimate, $\theta(k)$.
\subsection{The Extended Kalman Filter}
The EKF first predicts the mean and the minimum prediction MSE of the state ahead, $\hat \theta\left(k|k-1\right)$ and $P(k|k-1)$, respectively, given the previous values. Then, the EKF updates the estimates with the observation and computes the mean and the minimum MSE of a posteriori state estimate,  $\hat \theta(k|k)$ and $P(k|k)$, respectively. The Kalman gain, $K(k)$, indicates the amount of correction required for an observation sample. Since $z(\theta(k))$ is a nonlinear function, it is linearized with a first-order Taylor expansion. Therefore, $\dot z(k)$ denotes the Jacobian of $z(\theta(k))$ with respect to $\theta$.

The first and the second moments of the state ahead are predicted as
\begin{eqnarray}
            \hat \theta(k|k-1)&=&\hat \theta(k-1|k-1) \label{eqn:hatTheta1}\\
            P(k|k-1) &=& P(k-1|k-1)+\sigma^2_\Delta.
\end{eqnarray}
After the observation, the posteriori state estimate statistics are updated as
\begin{eqnarray}
            \hat \theta(k|k)&=&\hat \theta(k|k-1) + \Re \{ K(k)( y(k)-{z}(\hat \theta(k|k-1))) \}\\
            P(k|k) &=& (1-\Re\{ K(k) \dot z(k)\})P(k|k-1),
\end{eqnarray}
where the Kalman gain is determined as
\begin{eqnarray}
             K(k)&=&P(k|k-1)\dot z(k)^H\Big( C+\dot z(k) P(k|k-1)\dot z(k)^H \Big)^{-1}, \label{eqn:KalmanGain}
\end{eqnarray}
the Jacobian of $z$ with respect to $\theta$ is given by
\begin{eqnarray}
\dot z(k)&=&\frac{\partial {z}(\theta(k))}{\partial \theta(k)}\Big |_{\hat \theta(k|k-1)}=
j{z}(\hat \theta(k)).
\end{eqnarray}
In (\ref{eqn:KalmanGain}),  $ C$ is the observation noise variance, given by
\begin{eqnarray}
C&=& \frac{\sigma^2_w}{2} + j\frac{\sigma^2_w}{2} \label{eqn:Ck}.
\end{eqnarray}

The EKF needs to be properly initialized to compute $\hat \theta(1|0)$ and the MSE $P(1|0)$. It is assumed that a sufficient number of pilots is used for frequency, frame and phase synchronization at the beginning of the frame. Therefore, there is no phase shift of the first received signal i.e. $\hat \theta(1|0)=0$. Moreover, at the first time instant, the minimum prediction MSE is set to $P(1|0)=\sigma^2_\Delta$ since this error amounts to the estimation of $\hat \theta(0|0)$ without any data.

\begin{figure}[htbp!]
\begin{center}
\includegraphics[width=12cm]{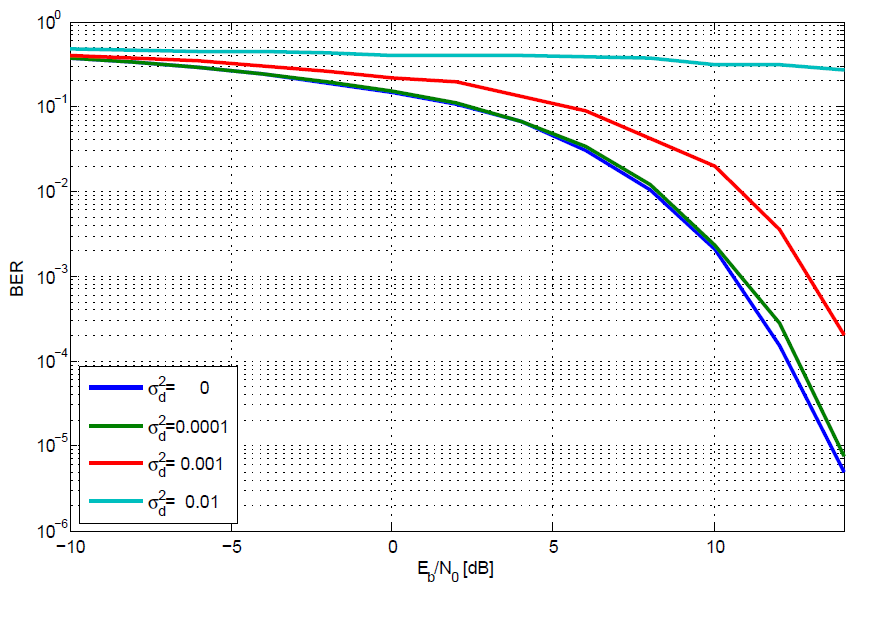}\center
\caption{BER vs. $E_b/N_0$ for 16-QAM for various values of $\sigma^2_\Delta$.}\label{fig:fig1}
\end{center}
\end{figure}

The EKF is applied for different phase noise innovation variance values. The bit error rate (BER) performance of the system is investigated. BER vs. $E_b/N_0$ for 16-QAM is shown in Fig. \ref{fig:fig1}. It can be seen that as $E_b/N_0$ increases, the EKF performs better and can track phase noise processes with higher variances.

\section{Coded SISO}\label{sec:CodedSISO}

In coded SISO systems, the  data bits are first encoded with an LDPC encoder. Then, the encoded bits are modulated onto complex symbols from an $M$-QAM constellation $\mathcal M$ and transmitted through the AWGN channel. 
 The phase noise process should be estimated in order to de-rotate the signal space.

\begin{figure}[t]
\begin{center}
\includegraphics[width=11cm]{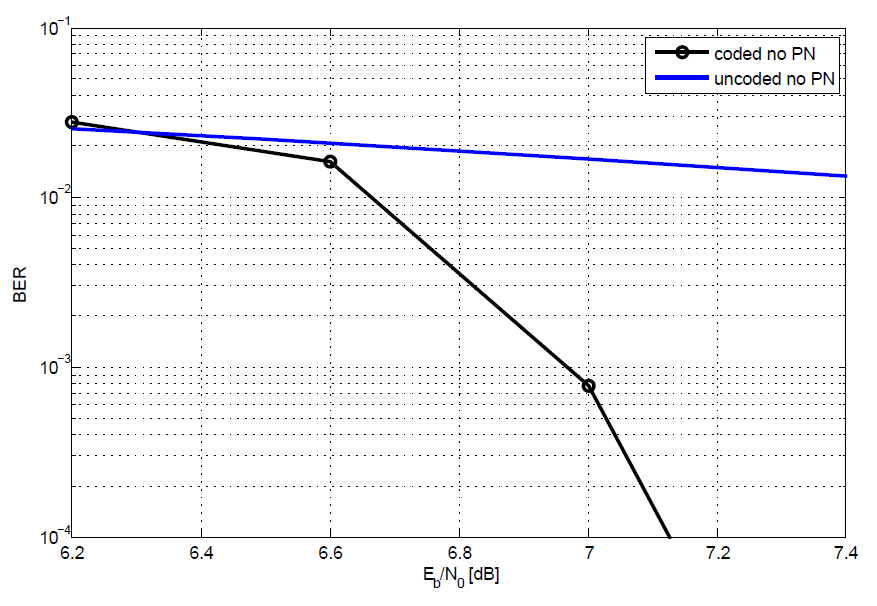}\center \caption{BER vs. $E_b/N_0$ for 16-QAM uncoded and coded system without phase noise}\label{fig:noPNcomparison}
\end{center}
\end{figure}

Channel coding adds redundant bits to protect information bits and to correct erroneous bits. A rate $R=7/8$ regular LDPC encoder  from the NASA Goddard technical standard \cite{Goddard} is used. It is a regular code with variable node degree 4 and check node degree 32. The BER vs. $E_b/N_0$ for both uncoded and coded systems without phase noise is shown in Fig. \ref{fig:noPNcomparison}. Data bits are mapped to 16-QAM symbols. As expected,  very low BERs can be achieved with channel coding at medium and high $E_b/N_0$ levels.

In a coded SISO system in the presence of phase noise, decoding and phase noise estimation can be performed jointly in an iterative way to complement one another. The EM framework is used in \cite{Herzet1} to estimate the constant phase offset. 
In \cite{Shehata}, a modified EM-based algorithm is suggested to estimate time varying phase noise. The tracking is performed by both an extended Kalman filter and smoother (EKS) and a linearized Kalman filter and smoother with  maximum-likelihood average (KS-MLA). The MLA algorithm removes the phase noise average over the frame. In \cite{Shehata}, the modified EM-based algorithm  is not derived analytically. In the following, we demonstrate analytically how an EM-based algorithm can be utilized for code-aided synchronization. The performance of the proposed estimators is investigated numerically.



First, recall that the signal model and the phase noise model for the uncoded SISO system is given by
\begin{eqnarray}
 \theta(k)&=&\theta(k-1)+\Delta(k)\nonumber\\
  y(k)&=&s(k) e^{j\theta(k)}+w(k)\nonumber.
\end{eqnarray}
Let $L_f$ denote the number of symbols in one frame. We define the vectors
\begin{eqnarray}
\mathbf y &\triangleq & [ y(1) ,  y(2) , \dots , y(L_f)]^T  \\
\mathbf s &\triangleq & [ s(1) ,  s(2) , \dots , s(L_f)]^T \\
\boldsymbol \theta &\triangleq & [ \theta(1) , \theta(2) , \dots , \theta(L_f)]^T .
\end{eqnarray}

Given the observation vector, $\mathbf y$, \emph{the maximum a posteriori} (MAP) estimate of $\boldsymbol \theta$ is given by
\begin{align}
\hat{ \boldsymbol \theta}&=\textrm{arg} \max_{\boldsymbol \theta} \ln p(\mathbf y| {\boldsymbol \theta})+\ln p({ \boldsymbol \theta})\label{eqn:MAPsisoCoded}
\end{align}
where $p(\mathbf y| {\boldsymbol \theta})=\displaystyle\sum_{\mathbf s} p(\mathbf s) p(\mathbf y|\mathbf s, {\boldsymbol \theta})$ is the a priori probability of parameters $\boldsymbol \theta$.

The EM framework is an iterative approach to solve an estimation problem where the observation, $\mathbf y$, depends not only on the parameter to be estimated, $\boldsymbol \theta$, but also on some nuisance parameter, $\mathbf s$. Note that transmitted symbols, $\mathbf s$, and phase noise $\boldsymbol \theta$ are independent. The EM algorithm iteratively maximizes the conditional a posteriori expectation, $E_{\mathbf s}\Big\{\ln p(\mathbf y|\mathbf s,\tilde{\boldsymbol \theta})|\mathbf y,\hat{\boldsymbol \theta}^{(i-1)}\Big\}+\ln p(\boldsymbol \theta)$, where $\ln f(\mathbf y|\mathbf s,\tilde{\boldsymbol \theta})$ is the conditional log likelihood function (LLF) and $E\{ . \}$ denotes the expectation operator. In the expectation step the block of previous phase noise estimates, $\hat{\boldsymbol \theta}^{(i-1)}$, is kept fixed and it is updated in the maximization step.

The EM framework for the $i$th iteration is defined by \cite{Herzet1}
\begin{eqnarray}
\textrm{Expectation step:} && Q\Big(\tilde{\boldsymbol \theta}| \hat{\boldsymbol \theta}^{(i-1)}\Big) \triangleq
 E_{\mathbf s}\Big\{\ln p(\mathbf y|\mathbf s,\tilde{\boldsymbol \theta})|\mathbf y,\hat{\boldsymbol \theta}^{(i-1)}\Big\}+\ln p(\boldsymbol \theta) \\
 \textrm{Maximization step:} && \hat{\boldsymbol \theta}^{(i)}=\arg\max_{\tilde{\boldsymbol \theta}}\Big\{Q\Big(\tilde{\boldsymbol \theta}| \hat{\boldsymbol \theta}^{(i-1)}\Big)\Big\}\label{eqn:MstepSISO}.
\end{eqnarray}
The EM algorithm has been shown to converge to the MAP solution if it is accurately initialized.

\subsection{Expectation Step (E-Step)}

Given the transmitted data and the phase noise process, the received signal has a Gaussian distribution. After taking logarithm and dropping constant terms, the conditional LLF is given by
\begin{eqnarray}
\ln p(\mathbf y|\mathbf s,\tilde{\boldsymbol \theta})\varpropto -2\Re \sum_{k=1}^{L_f}y(k) s^{*}(k)e^{-j\theta(k)}. \label{eqn:lnEstepSISO}
\end{eqnarray}
Taking the conditional expectation over $\mathbf s$, given $\mathbf y$ and $\hat{\boldsymbol \theta}^{(i-1)}$, equation for the expectation step can be written as
\begin{eqnarray}
Q\Big(\tilde{\boldsymbol \theta}| \hat{\boldsymbol \theta}^{(i-1)}\Big)= -2\Re \sum_{k=1}^{L_f} \alpha(k)^*y(k)e^{j\theta(k)}+\ln p({ \boldsymbol \theta}) \label{eqn:qFunc},
\end{eqnarray}
where the weighted or soft symbol $\alpha(k)$ is defined as
\begin{eqnarray}
\alpha(k) \triangleq \sum_{a_l\in \mathcal M} a_l p\left(s(k)=a_l|\mathbf y, \hat{\boldsymbol \theta}^{(i-1)}\right) \label{eqn:alpha1}.
\end{eqnarray}
In (\ref{eqn:alpha1}),$p(s(k)=a_l|\mathbf y, \hat{\boldsymbol \theta}^{(i-1)})$ are the marginal a posteriori symbol probabilities (APPs) of the transmitted symbol. For 16-QAM, if Gray mapping  is used, (\ref{eqn:alpha1}) can be computed as
\begin{eqnarray}
\alpha(k) = \beta_1(k)(2-\beta_2(k))-j\beta_3(k)(2-\beta_4(k)) \label{eqn:alpha},
\end{eqnarray}
where
\begin{eqnarray}
\beta_i(k)=\tanh \frac{L_i(k)}{2}
\end{eqnarray}
and $L_i(k)$ is the log likelihood ratio (LLR) of the a posteriori probability of the $i$ th bit of the symbol at  time instant $k$, $b_i(k)$. The LLR is defined as
\begin{eqnarray}
L_i(k)=\log \frac{p(b_i(k)=1|\mathbf y, \hat{\boldsymbol \theta}^{(i-1)})}{p(b_i(k)=0|\mathbf y, \hat{\boldsymbol \theta}^{(i-1)})}.
\end{eqnarray}

\begin{figure}[t]
\begin{center}
\includegraphics[width=16cm]{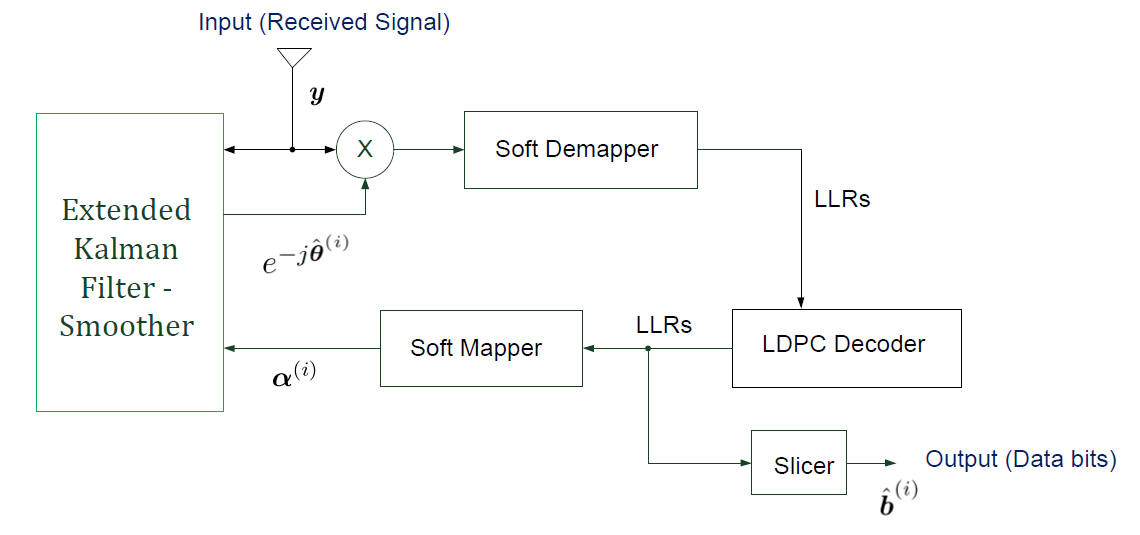}\center \caption{Block diagram of the receiver structure for coded SISO system.}\label{fig:figRS_SISO}
\end{center}
\end{figure}

The block diagram of the receiver structure is shown in Fig. \ref{fig:figRS_SISO} where superscript $(i)$ denotes the iteration of the EM algorithm. The LLRs are first computed by a soft demapper (demodulator). Then, they are passed to the LDPC decoder. Next, the LDPC decoder runs a sufficient number of iterations to achieve more accurate LLRs. After that, the LLRs computed by the LDPC decoder are modulated into soft symbols according to (\ref{eqn:alpha}) by a soft mapper. As a result, the expectation step computes the block of APPs denoted as
\begin{eqnarray}
\boldsymbol \alpha &\triangleq & [ \alpha(1) , \alpha(2) , \dots , \alpha(L_f)]^T .
\end{eqnarray}

\subsection{Maximization Step (M-Step)}
The maximization step maximizes $Q\Big(\tilde{\boldsymbol \theta}| \hat{\boldsymbol \theta}^{(i-1)}\Big)$ in (\ref{eqn:MstepSISO}) with respect to $\boldsymbol \theta$.
 Since the system can be modeled in a state-space form, the solution can be given by Kalman filtering. The Kalman filter is an optimal MMSE estimator which can be considered as a MAP estimator \cite{SMKay}. Then, a decision directed Kalman filter where the transmitted symbol vector, $\mathbf s$, is replaced by the soft decision symbol vector, $\boldsymbol \alpha$, estimates $\boldsymbol \theta$ as
\begin{align}
\hat{ \boldsymbol \theta}= \textrm{arg}\, \max_{ { \boldsymbol \theta}} \ln\ p(\mathbf y| {\boldsymbol \theta}, \mathbf s=\boldsymbol \alpha)+\ln p({ \boldsymbol \theta}).\label{eqn:MAPthetaSISO}
\end{align}

By setting $\mathbf{s}= \boldsymbol \alpha$, the conditional LLF in (\ref{eqn:lnEstepSISO}) can be rewritten as
\begin{align}
\ln p(\mathbf s|\mathbf s=\boldsymbol \alpha, {\boldsymbol \theta})\varpropto 2\Re \sum_{k=1}^{L_f} y(k)  \alpha^*(k) e^{-j\theta(k)}\label{eqn:lnEstepSISO2}.
\end{align}
Using (\ref{eqn:lnEstepSISO2}) in (\ref{eqn:MAPthetaSISO}) we achieve (\ref{eqn:qFunc}) which is the function to be maximized in the M-step. As a result, a decision directed Kalman filter can be used to carry the M-step.

 In \cite{Shehata}, two soft decision-directed Kalman filters, the KS-MLA and the EKS, are proposed. In the following, two estimators will be explained.

\subsubsection{The Kalman Filter and Smoother with Maximum Likelihood Average}
The KS-MLA proposes a new observation equation which is a linear function of $\theta$. Assuming that $\alpha(k)=s(k)$, the KS-MLA first removes the data dependency from the measurement via
\begin{eqnarray}
 \epsilon(k)&=&Im\{y(k)\alpha^*(k)\}
\end{eqnarray}
where $\epsilon(k)$ is the new observation equation. Then, the KS-MLA assumes that the phase noise values are small enough such that
\begin{eqnarray}
\sin \theta &\approx& \theta \label{eqn:sintheta}\\
 \cos \theta &\approx& 1. \label{eqn:costheta}
\end{eqnarray}
 Subsequently, new observation equation is given by
\begin{eqnarray}
 \epsilon(k)&=&\Im\{y(k)\}\Re\{\alpha(k)\}-\Re\{y(k)\}\Im\{\alpha(k)\}\\
 &=&u(k)\theta(k)+v(k)\label{eqn:newOBS},
\end{eqnarray}
where
\begin{eqnarray}
 u(k)&=&\Re\{s(k)\}\Re\{\alpha(k)\}+\Im\{s(k)\}\Im\{\alpha(k)\}\\
 &=&\parallel \alpha(k)\parallel^2,
\end{eqnarray}
and
\begin{eqnarray}
 v(k)&=&\Re\{\alpha(k)\}\Im\{w(k)\}+\Im\{\alpha(k)\}\Re\{w(k)\},
\end{eqnarray}
is a zero mean Gaussian random variable with variance $\sigma^2_v=E_s\sigma^2_w/2$ and $E_s$ is the average symbol energy. Since the observation equation is linear, a regular Kalman filter and smoother (KS) can be used. Note that the new observation equation does not depend on the complex transmitted symbol. Instead, it is a function of the amplitude square of the soft symbol.

 Phase noise process reaches large values within a block. Depending on $\sigma^2_\Delta$ and $L_f$, it has a nonzero average over the block. The MLA algorithm estimates the ML average of the phase noise process over each block and removes it from each received block, $\mathbf y$, at the input of KS. Then, the parameter to be estimated is given by
\begin{eqnarray}
 \psi(k)&=&\theta(k)-\theta_{avg}
\end{eqnarray}
where
\begin{eqnarray}
 \theta_{avg}&=&\textrm{arg}\sum_{k=1}^{N_s} y(k)\alpha(k)^* .
\end{eqnarray}
The MLA algorithm is directly embedded into the EM framework and performed before the estimator in the M-step.

\subsubsection{Extended Kalman Filter and Smoother}

Second estimator proposed in \cite{Shehata} is the soft decision-directed EKS. The observation equation in (\ref{eqn:obsUncoded}) is a nonlinear function of $\theta$. Therefore, a suboptimal decision directed EKS is used to track the phase noise process in \cite{Shehata}.
The state and observation equations at time $k$ are given as
\begin{eqnarray}
\textrm{State:} && \theta(k)=\theta(k-1)+\Delta(k)\\
\textrm{Observation:} && y(k)=\alpha(k) e^{j\theta(k)}+w(k) \label{eqn:obscoded}.
\end{eqnarray}
Note that (\ref{eqn:obsUncoded}) and (\ref{eqn:obscoded}) are  similar. We can rewrite the observation equation as
\begin{eqnarray}
y(k)&=& z_c(\theta(k))+w(k),
\end{eqnarray}
where $z_c$ stands for the nonlinear function for the coded system. The EKS approximates $z_c$ as
\begin{eqnarray}
z_c(\theta(k))\approx z_c(\hat \theta(k|k-1)) + \frac{\partial {z_c}}{\partial \theta(k)}\Big |_{\hat \theta(k|k-1)}(\theta(k)-\hat \theta(k|k-1)) \label{eqn:thetaTaylor},
\end{eqnarray}
where $\hat \theta(k|k-1)$ is the estimate of $\theta(k)$ based on the previous data.

The EKS first filters the received signal and estimates the block of parameters. Afterwards, it smooths the estimations with a backward recursion. It is initialized with state estimate $\hat \theta(1|0)=0$ and a posteriori MSE $P(1|0)=\sigma^2_\Delta$.
The filtering equations compute $\hat \theta(k)$ and $P(k),k=2,3,\dots,N_s$ in a recursive fashion according to (\ref{eqn:hatTheta1}-\ref{eqn:Ck}) where $z(\theta(k))$ is replaced by $z_c(\theta(k))$.

The smoother runs a backward recursion to find better estimates of the parameter block. We will use subscript $s$ to refer to smoothing. $k^+$ denotes the sequence from $k$ to $L_f$. The set of equations for smoothing are given by
\begin{eqnarray}
\hat \theta_s(k|k^+)&=&\hat \theta(k|k) + P(k|k) P(k+1|k)^{-1}(\hat \theta_s(k+1|k+1^+)-\hat \theta(k|k)) \\
P_s(k|k^+)&=&P(k|k)+P(k|k)P(k+1|k)^{-1}\nonumber\\
&&\qquad \times(P_s(k+1|k+1^+)-P(k+1|k))(P(k|k)P(k+1|k)^{-1})^T
\end{eqnarray}

\subsection{LDPC decoder}

LDPC codes are linear block codes that have a sparse parity check matrix $\mathbf H_p$. The input of the LDPC decoder are the LLRs that are computed by the soft demapper (demodulator). The output of the LDPC decoder are the updated LLRs that will be used by the soft mapper (modulator).

\begin{figure}[H]
\begin{center}
\includegraphics[width=11cm]{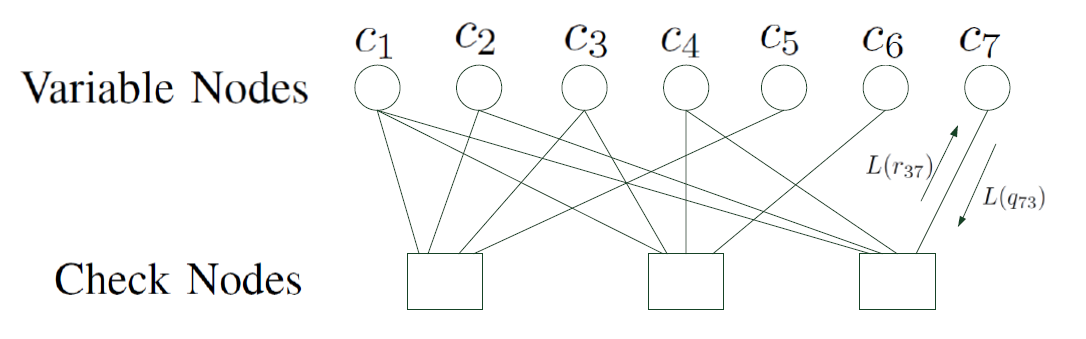}\center \caption{Tanner graph for $\mathbf H_p$ in (\ref{eqn:Hldpc}).}\label{fig:Tanner1}
\end{center}
\end{figure}

An example of a  parity check matrix of a (7,4) irregular block code, i.e. the number of 1's in each row and in each column is not constant, is given by
\begin{eqnarray}
\mathbf H_p=\left[ \begin{array}{ccccccc}
1&1&1&0&1&0&0 \\
1&0&1&1&0&1&0 \\
1&1&0&1&0&0&1 \end{array} \right].\label{eqn:Hldpc}
\end{eqnarray}
The corresponding Tanner graph is shown in Fig. \ref{fig:Tanner1}. There are 7 variable nodes and 4 check nodes for $\mathbf H_p$. Let $k=1,2,3$ an $l=1,2,\dots,7$ be the index of the rows and columns of $\mathbf H_p$, respectively. If $\mathbf H_p$[k,l] is 1, then there exists a connection between the check node $k$ and the variable node $l$. Note that the information exchange is done in two ways for each connection, one from variable node to check node and one for the check node to variable node.

The LDPC decoder runs the sum product algorithm to estimate the LLRs of the transmitted bits. It computes the LLR of each path from variable nodes to check nodes, and from check nodes to variable nodes iteratively. $L(c_l)$ denotes the LLR of the $l$th bit of the transmitted codeword $c$. Let $L(r_{kl})$ denote the LLR that belongs to the connection in the direction of check node $k$ to variable node $l$. Similarly, $L(q_{lk})$ denotes the LLR of the same connection in the opposite direction. An example is also shown in Fig. \ref{fig:Tanner1}. $L(Q_l)$ denotes the updated LLR of the $l$th bit of the transmitted codeword. The sum product algorithm is performed as follows
\begin{eqnarray}
L(r_{kl})&=&2 \textrm{atanh} \prod_{l'\in V_{k\textbackslash l}}\tanh \Big(\frac{1}{2}L(q_{l'k})\Big) \label{eqn:Lr} ,\\
L(q_{lk})&=&L(c_l)+\sum_{k'\in C_{l\textbackslash k}}L(r_{k'l}), \\
L(Q_l)&=&L(c_l)+\sum_{k'\in C_l}L(r_{k'l}),
\end{eqnarray}
where $V_{k\textbackslash l}$ represents the set of indexes of all the variable nodes connected to the check node $k$ except for the variable node $l$. Similarly, $C_{l\textbackslash k}$ represents the set of indexes of all the check nodes connected to the variable node $l$ except for the check node $k$. For example, if $k=2$ and $l=4$, $V_{2\textbackslash 4}=\{1,3,6\}$, $C_4=\{2,3\}$ and $C_{4\textbackslash 2}=\{3\}$.

\subsubsection{Improving  the speed of the algorithm}

$L(q_{lk})$ should be initialized for the first iteration. Normally, it is initialized with the input LLRs as
\begin{eqnarray}
L(q_{lk})=L(c_{l}).
\end{eqnarray}

As suggested in \cite{Herzet1}, instead of initializing with the input LLRs, $L(q_{lk})$ is initialized by the last update of it from the latest EM algorithm iteration. Thus,
\begin{eqnarray}
L(q_{lk})=L(q_{lk})^{(i-1)}
\end{eqnarray}
where $(i)$ denotes the algorithm iteration.
It is seen that significant complexity reduction can be achieved with negligible degradation in BER with this approximate version. The number of the decoder iterations can be reduced and the number of the EM algorithm iterations can be increased. Since the complexity heavily depends on the number of decoder iterations, the overall complexity is much less than that of the original EM-based algorithm.


\subsection{Simulation Results}

\begin{figure}[t]
\begin{center}
\includegraphics[width=11cm]{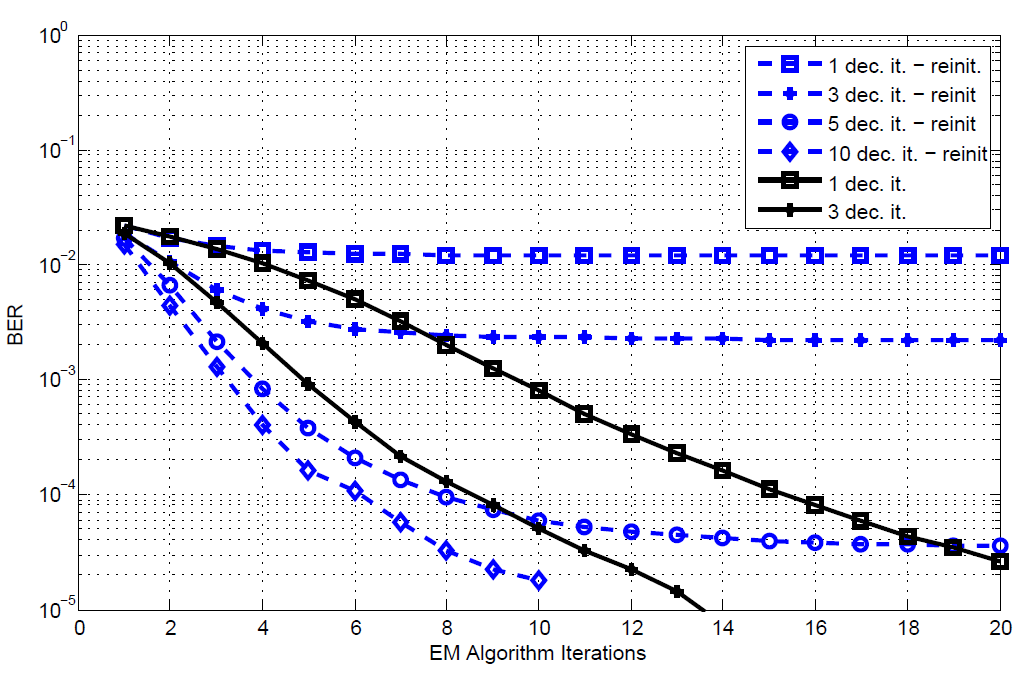}\center \caption{BER vs. the EM algorithm iterations for 256-QAM coded system with EKS at $E_b/N_0=16$dB for different number of decoder iterations, with and without keeping internal decoder information.} \label{fig:decode_no_restart_256}
\end{center}
\end{figure}

First, the advantage of keeping internal information at the decoder is investigated for the 256-QAM system with the EKS. In Fig. \ref{fig:decode_no_restart_256}, the BER is calculated for each EM iteration for different number of decoding iterations with and without keeping internal information at the decoder for a fixed $E_b/N_0=16$dB. The dashed lines represent the algorithm where the LDPC decoder is reinitialized between each iteration of the EM algorithm. In Fig. \ref{fig:decode_no_restart_256}, the solid lines represent the new algorithm where the internal information in the LDPC decoder is kept between the EM algorithm iterations. It is seen that with keeping internal information overall complexity can be reduced due to fewer iterations. When the LDPC decoder is reinitialized the BER tends to reach a floor and the system cannot perform under this level even with more EM iterations. On the other hand, when we keep the internal information at the LDPC decoder, lower BERs can be achieved with increasing number of EM iterations since better LLRs are obtained from the decoder in each EM iteration.

\begin{figure}[t]
\begin{center}
\includegraphics[width=11cm]{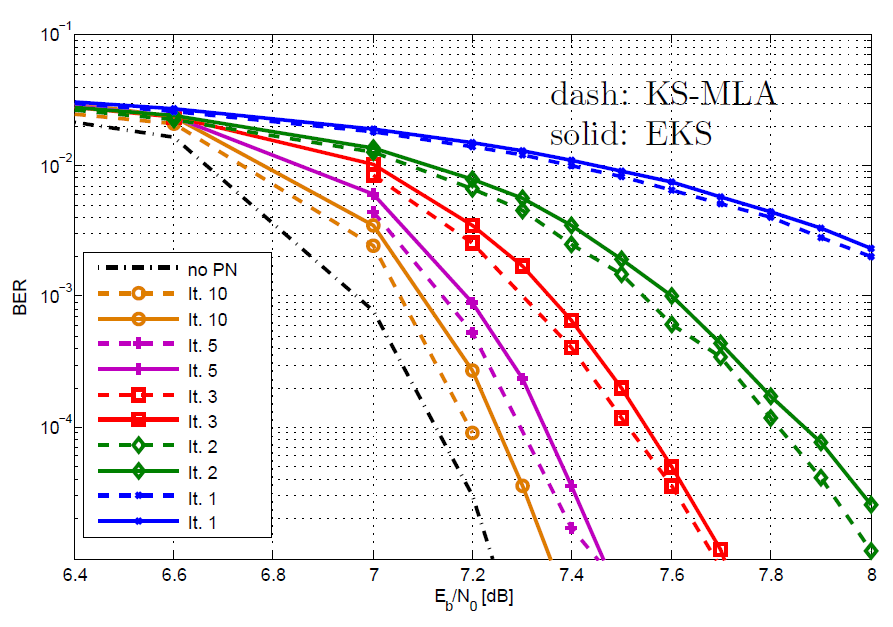}\center \caption{BER vs. $E_b/N_0$ for 16-QAM coded system with both the EKS and the KS-MLA where $\sigma^2_\Delta=10^{-4}$ and 3 decoding iteration.} \label{fig:BERvsSNR1e43dec}
\end{center}
\end{figure}

%
%
%
%
%
In Fig. \ref{fig:BERvsSNR1e43dec}, the results of the EM algorithm are shown for different number of iterations for both the EKS and the KS-MLA for a 16-QAM system where $\sigma^2_\Delta=10^{-4}$. The number of iterations performed by the LDPC decoder is set to 3. It is seen that the KS-MLA performs slightly better than the EKS.

\begin{figure}[t]
\begin{center}
\includegraphics[width=11cm]{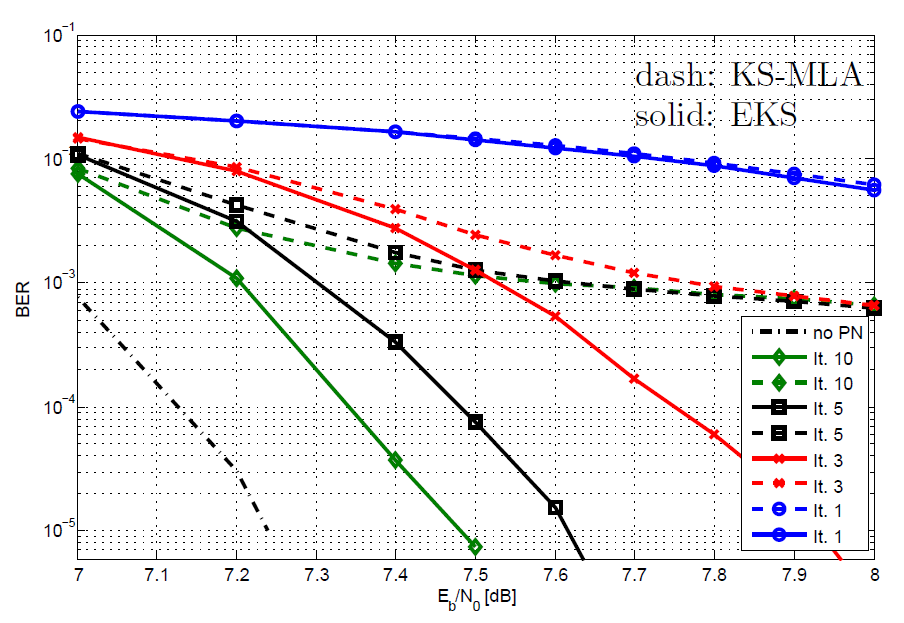}\center \caption{BER vs. $E_b/N_0$ for 16-QAM coded system with both the EKS and the KS-MLA where $\sigma^2_\Delta=3 \cdot 10^{-4}$ and 3 decoding iteration.} \label{fig:BERvsSNR3e43dec}
\end{center}
\end{figure}

The KS-MLA suggests a new linear observation equation (\ref{eqn:newOBS}) by assuming $\alpha(k) = s(k)$. Moreover, the received signal is multiplied by the soft decision symbol which also increases the noise power. This \emph{ad hoc} method is proposed in \cite{Shehata} to alleviate the effects of the assumptions in (\ref{eqn:sintheta}) and (\ref{eqn:costheta}). When the phase noise values are small, $\alpha(k)$ is a reliable estimate of $s(k)$. Then, the KS-MLA algorithm performs  better than  the EKS.
However, the KS-MLA fails to track more severe phase noise processes. In Fig.
\ref{fig:BERvsSNR3e43dec}, the phase noise innovation variance is set to $\sigma^2_\Delta=3 \cdot 10^{-4}$ where all the other parameters remain unchanged. It is concluded that the assumptions (\ref{eqn:sintheta}) and (\ref{eqn:costheta}) are violated and $\alpha(k)$ is a less reliable estimate of $s(k)$. Thus, the KFS-MLA performs significantly worse than the EKS. Moreover, the KS-MLA has an irreducible error floor.

\begin{figure}[t]
\begin{center}
\includegraphics[width=11cm]{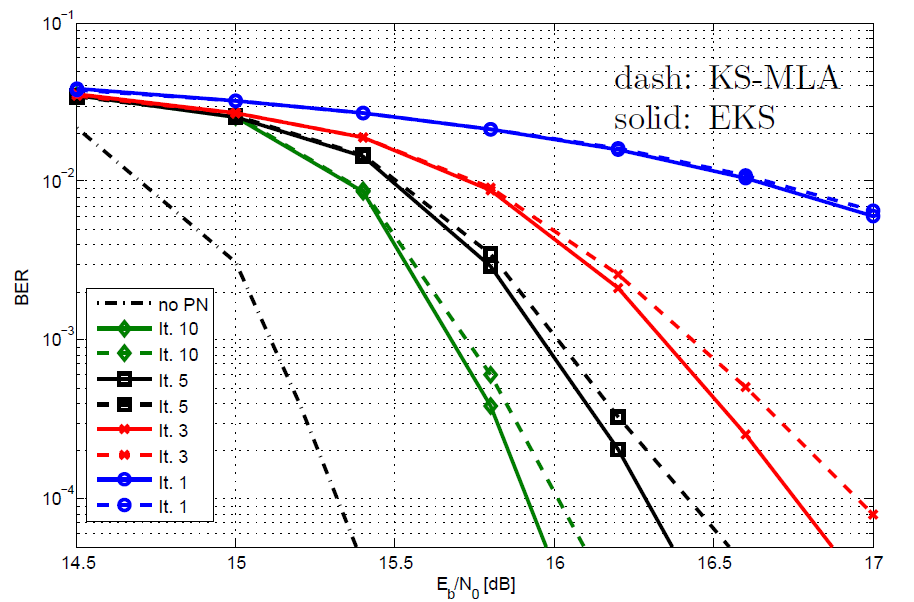}\center \caption{BER vs. $E_b/N_0$ for 256-QAM coded system with both the EKS and the KS-MLA where $\sigma^2_d=10^{-4}$ and 3 decoding iteration.} \label{fig:BERvsSNR1e43dec_256}
\end{center}
\end{figure}

In Fig. \ref{fig:BERvsSNR1e43dec_256}, the results for both the EKS and the KS-MLA are shown for a 256-QAM system where $\sigma^2_\Delta=10^{-4}$. It is well-known that the reliability of the estimate of a 256-QAM symbol is less than that of a 16-QAM symbol for the same phase noise process. The reason is that phase noise effects the outer symbols in the signal constellation. In addition, the 256-QAM constellation is more densely packed than the 16-QAM constellation, i.e., the distance and the phase difference between two neighbor constellation points is smaller. Therefore, the performance of the KS-MLA is degraded since the noise power is also amplified. In Fig. \ref{fig:BERvsSNR1e43dec_256}, we observe that the EKS performs better than the KS-MLA. As a result, the EKS is more advantageous than the KS-MLA for higher order modulations.

\chapter{Phase Noise Estimation \\ for MIMO systems}\label{chap:MIMO-model}
The demand for high data rate wireless communications  has spearheaded  the research in the field of MIMO systems. It is shown that the bandwidth efficiency of a wireless link can be significantly improved with the usage of MIMO systems \cite{Telatar99,Foschini98}. However, synchronization errors decrease the performance of MIMO system dramatically. In Sec. \ref{sec:UncodedMIMO} the system model for an $N_t\times N_r$ uncoded MIMO systems is introduced. Detection in MIMO systems is also briefly discussed. An EKF is applied to track multiple phase noise processes. In Sec. \ref{sec:CodedMIMO}, the main contribution of this thesis is provided. Joint phase noise estimation and detection in coded MIMO systems is performed with the help of an EM-based algorithm for the first time in the literature. Exploiting LDPC codes, EKFS and BICM, the proposed EM-based algorithm iteratively solves the problem of joint phase noise estimation and detection. Simulation results are also evaluated for a $2\times2$ LOS-MIMO system over Rician fading MIMO channels.

\section{Uncoded MIMO}\label{sec:UncodedMIMO}
An uncoded MIMO system with $N_t$ transmit and $N_r$ receive antennas is under consideration.
At the transmitter, a group of data bits are modulated onto an $M$-QAM constellation $ \Omega$. They are demultiplexed into substreams of symbols of length $L_f$. Subsequently, using spatial multiplexing the symbols are transmitted simultaneously from $N_t$ antennas. Quasi-static block fading channels are considered, i.e., the channel gains remain constant over the length of a frame but change from frame to frame. Similar to previous work in the literature \cite{Hadaschik}, it is assumed that the $N_r\times N_t$ channel matrix $\mathbf{H}$ is estimated using orthogonal training sequences that are transmitted at the beginning of each frame. To ensure that the proposed scheme is applicable to LoS and multi-user MIMO systems, it is assumed that independent oscillators are deployed at each transmit and receive antenna.
AWGN is also taken into consideration. The received signal is also effected by time varying phase noise both at the transmitter and the receiver side. To ensure that the proposed scheme is applicable to LoS and multi-user MIMO systems, it is assumed that independent oscillators are deployed at each transmit and receive antenna.

The performance of the uncoded MIMO system is severely degraded in the presence of phase noise.  In \cite{Hadaschik}, pilot-aided estimation of phase noise in
an uncoded MIMO system is investigated. The phase noise parameters
corresponding to the transmit and receive antennas are estimated by applying a Wiener filtering approach. However, the proposed scheme is bandwidth inefficient and significant overhead is introduced. In \cite{Mehrpouyan}, joint channel and phase noise estimation is performed. A data aided least square (LS) estimator, a decision directed weighted least squares (WLS) estimator and a decision directed EKF are proposed to track time varying phase noise.

Based on the above set of assumptions, the received signal at time instance $k$, $\mathbf y(k) \triangleq \left[ y_1(k), y_2(k), \dots, y_{N_r}(k)  \right]^T$, is given by
\begin{align}\label{ref:obs1}
 \mathbf y(k) =& \boldsymbol \Gamma^{[r]}(k) \mathbf H \boldsymbol \Gamma^{[t]}(k) \mathbf s(k) + \mathbf w(k)
\end{align}
where
\begin{itemize}
\item $\boldsymbol \Gamma^{[r]}(k)\triangleq \textrm{diag}\Big(e^{j\theta_1^{[r]}(k)},e^{j\theta_2^{[r]}(k)},\dots,e^{j\theta_{N_r}^{[r]}(k)} \Big)$ is an $N_r\times N_r$ diagonal matrix and $\boldsymbol \Gamma^{[t]}(k) \triangleq \textrm{diag}\Big(e^{j\theta_1^{[t]}(k)},e^{j\theta_2^{[t]}(k)},\dots,e^{j\theta_{N_t}^{[t]}(k)} \Big)$ is an $N_t\times N_t$ diagonal matrix,
\item $\theta_\ell^{[r]}(k)$ and $\theta_m^{[t]}(k)$ denote the phase noise process at the $\ell$th receive and $m$th transmit antennas, respectively, where the $N_r\times N_t$ MIMO channel matrix is denoted as $\mathbf H \triangleq \left[ \mathbf h_1, \mathbf h_2, \dots, \mathbf h_{N_t}  \right]$ with $\mathbf h_{\ell} \triangleq \left[ h_{\ell,1},  h_{\ell,2}, \dots,  h_{\ell,N_t}  \right]^T$
\item $h_{\ell,m}$, for $\ell=1,2,\dots,N_r$ and $m=1,2,\dots,N_t$, denotes the channel gain between transmit antenna $m$ and receive antenna $\ell$ that is modeled as a complex Gaussian random variable, $\mathcal{CN}(\mu_{h_{m,\ell}}, \sigma^2_{h_{m,\ell}})$
\item $\mathbf s(k) \triangleq\left[ s_1(k), s_2(k), \dots, s_{N_t}(k)  \right]^T$ is the vector of transmitted symbols,
\item $\mathbf w(k) \triangleq\left[ w_1(k), w_2(k), \dots, w_{N_r}(k)  \right]^T$ is the vector of zero-mean additive white Gaussian noise (AWGN), $\mathcal{CN}(0, \sigma^2_{w_m})$.
\end{itemize}
Phase noise is modeled as discrete-time Wiener process. Therefore, for $\ell=1,2,\dots,N_r$ and $m=1,2,\dots,N_t$ \cite{Demir}
\begin{align}\label{eq:pn_model}
\theta_\ell^{[r]}(k)=&\theta_\ell^{[r]}(k-1)+\Delta_\ell^{[r]}(k)\notag \\
\theta_m^{[t]}(k)=&\theta_m^{[t]}(k-1)+\Delta_m^{[t]}(k),
\end{align}
where $\Delta_\ell^{[r]}(k)$ and  $\Delta_m^{[t]}(k)$ are the phase innovations for the $\ell$th receiver and $m$th transmit antenna, respectively. They are assumed to be white real Gaussian processes with $\Delta_\ell^{[r]}(k)\thicksim \mathcal N(0,\sigma^2_{\Delta_\ell^{[r]}})$ and $\Delta_m^{[t]}(k)\thicksim \mathcal N(0,\sigma^2_{\Delta_m^{[t]}})$.
Let us define $\boldsymbol \theta(k)$ as
\begin{align}
\boldsymbol \theta(k)\triangleq\left[ \theta_1^{[r]}(k), \dots, \theta_{N_r}^{[r]}(k), \theta_1^{[t]}(k), \dots, \theta_{N_t}^{[t]}(k)  \right]^T.
\end{align}
Then we have
\begin{align}\label{eqn:stateMIMO1}
  \boldsymbol{\theta}(k)=&\boldsymbol{\theta}(k-1)+\boldsymbol{\Delta}(k)
\end{align}
where $\boldsymbol{\Delta}(k)\triangleq\left[ \Delta_1^{[r]}(k), \dots, \Delta_{N_r}^{[r]}(k), \Delta_1^{[t]}(k), \dots, \Delta_{N_t}^{[t]}(k)\right]^T$.

\subsection{MIMO detection}
We want to detect the transmitted symbol vector $\mathbf s(k)$ in the maximum-likelihood (ML) sense.
For an $M$-QAM modulated $N_t\times N_r$ MIMO system, ML detection of  the transmitted symbol vector at time instant $k$ is given by
\begin{eqnarray}
\hat{\mathbf s(k)}=\arg\max_{\mathbf s \in \Omega^{N_t}}  \parallel \mathbf y- \mathbf f(\mathbf s, \hat{\boldsymbol \theta}(k),\mathbf H) \parallel \label{eqn:uncMIMOdetection}
\end{eqnarray}
where
\begin{eqnarray}
\mathbf f(\mathbf s, \hat{\boldsymbol \theta}(k),\mathbf H)\triangleq\boldsymbol \Gamma^{[r]}(k) \mathbf H \boldsymbol \Gamma^{[t]}(k) \mathbf s.
\end{eqnarray}
The ML detection search over $M^{N_t}$ combinations, $S_i,i=1,2,\dots,M^{N_t}$, of the transmitted symbol vector. The problem in (\ref{eqn:uncMIMOdetection}) is  computationally infeasible especially for higher order modulations. There are several detectors that provide approximate solutions with low complexity, such as zero-forcing detection with decision feedback \cite{Golden,Wolniansky}, sphere detection \cite{Viterbo,Barbero}, and lattice reduction aided detection \cite{Agrell,Windpassinger}. However, low complexity MIMO detection algorithms are beyond the scope of this thesis. The ML detection problem in (\ref{eqn:uncMIMOdetection}) is directly solved by searching over  all possible transmitted symbol vectors.

\subsection{The Extended Kalman Filter}

The ML detector provides hard decision symbol vector, $\hat {\mathbf s}(k)$. Next, $\hat {\mathbf s}(k)$ is input to the EKF which estimates phase noise vector, $\boldsymbol \theta(k)$. A hard decision directed EKF is applied to estimate phase noise parameters.
Let us define $\mathbf X(k) \triangleq   \boldsymbol \Gamma^{[r]}(k) \mathbf H \boldsymbol \Gamma^{[t]}(k)$. Consequently, the vector of received signals at the receive antennas at time $k$, $\mathbf y(k)$ can be rewritten as
\begin{eqnarray}
 \mathbf y(k) &=&  \mathbf X(k) \mathbf s(k) + \mathbf w(k)\label{eq:received_signal2}\\
 &\approx& \mathbf X(k) \hat {\mathbf s}(k) + \mathbf w(k)\label{eq:obsMIMOunc2}.
\end{eqnarray}

It can be seen from (\ref{eqn:stateMIMO1}) and (\ref{eq:obsMIMOunc2}) that there are a total of $N_r+N_t$ phase noise processes that need to be tracked. However, as shown below, an equivalent signal model can be obtained that reduces the dimensionality of the estimation problem and results in reduces overhead and improved estimation accuracy \cite{Mehrpouyan}.

By arbitrarily selecting the phase noise process $\theta_{N_t}^{[t]}(k)$ as a reference phase noise value, $\mathbf X(k)$ in (\ref{eq:received_signal2}) can be rewritten as
\begin{align}
\mathbf X(k)
&=e^{j\left(\theta_{N_t}^{[t]}(k)\right)}{\boldsymbol \Gamma}^{[r]}(k) \mathbf H {\boldsymbol \Gamma}^{[t]}(k)e^{-j\left(\theta_{N_t}^{[t]}(k)\right)} \notag\\
&=\tilde {\boldsymbol \Gamma}^{[r]}(k) \mathbf H \tilde {\boldsymbol \Gamma}^{[t]}(k),
\end{align}
where
\begin{itemize}
\item$\tilde{\boldsymbol \Gamma}^{[r]}(k)\triangleq \textrm{diag}\Big\{e^{j(\phi_1(k))}, \dots, e^{j(\phi_{N_r}(k))}$ \Big\},
\item $\tilde{\boldsymbol \Gamma}^{[t]}(k)\triangleq \textrm{diag}\Big\{e^{j(\phi_{N_r+1}(k))}, \dots, e^{j(\phi_{N_r+N_t-1}(k))}, 1 \Big\}$,
\item $\boldsymbol{\phi}(k) \triangleq \left[ \phi_1(k),\cdots,\phi_{N_r+N_t-1}(k)\right]^T$, and
        \begin{align*}
        \phi_f(k)\triangleq
        \begin{cases}
            \theta^{[r]}_f+\theta^{[t]}_{N_t}&f=1,\cdots,N_r,\\
            \theta^{[t]}_{f-N_r}-\theta^{[t]}_{N_t}&f=N_r+1,\cdots,N_r+N_t-1.
        \end{cases}
       \end{align*}
\end{itemize}

It is seen that $N_r+N_t$ state parameters can be replaced to $N_r+N_t-1$ new state parameters without changing the observation equalities. Then, we can rewrite the state and observation equations at time $k$ as
\begin{eqnarray}
\boldsymbol{\phi}(k)&=&\boldsymbol{\phi}(k-1)+\boldsymbol{\Delta}(k)\label{eqn:state4}\\
\mathbf y(k) &=& \mathbf z(\boldsymbol{\phi}(k))+ \mathbf w(k)\label{eqn:obsMIMOunc42},
\end{eqnarray}
 where
\begin{eqnarray}
\mathbf z(\boldsymbol{\phi}(k))&\triangleq&\left[ z_1(\boldsymbol{\phi}(k)), z_2(\boldsymbol{\phi}(k)), \dots,  z_{N_r}(\boldsymbol{\phi}(k)) \right]^T,\nonumber\\
 &=&\tilde {\boldsymbol \Gamma}^{[r]}(k) \mathbf H \tilde {\boldsymbol \Gamma}^{[t]}(k) \hat {\mathbf s}(k),\nonumber
\end{eqnarray}
and $z_m(\boldsymbol{\phi}(k)), m=1,2,\dots,N_r$, denotes the $m$th element of vector $\mathbf z(\boldsymbol{\phi}(k))$. Note that we assume that $\mathbf H$ is known, $\hat {\mathbf s}(k)$ is provided by the detector and $z_m(\boldsymbol{\phi}(k))$ is a nonlinear function of $\boldsymbol{\phi}(k)$.

It is assumed that perfect phase synchronization is obtained at the beginning of a frame by transmitting sufficient number of pilots.  Then, the EKF is initialized with the state estimate $\hat {\boldsymbol \phi}(0)=0$ and the error covariance estimate $\hat {\mathbf M}(0)=\sigma^2_\Delta\mathbf I$.  The EKF first estimates the {a priori} state vector $\hat {\boldsymbol \phi}^-(k)$,  and the $(N_r+N_t-1)\times (N_r+N_t-1)$ {a priori} error covariance matrix $\hat {\mathbf M}^{^-}(k)$ as follows
\begin{eqnarray}
            \hat {\boldsymbol \phi}^-(k)&=&\hat {\boldsymbol \phi}(k-1) \label{eqn:hatPhiMIMO}\\
            \hat {\mathbf M}^{^-}(k) &=& \hat {\mathbf M}(k)+ 2 \sigma^2_\Delta \mathbf I.
\end{eqnarray}
Next, the EKF linearizes  the nonlinear function $\mathbf z(\boldsymbol{\phi}(k))$ in (\ref{eqn:obsMIMOunc42})  about the a priori estimate of the state vector using a first-order Taylor approximation as
\begin{eqnarray}
\mathbf z( {\boldsymbol \phi}(k))&\approx& \mathbf z(\hat {\boldsymbol \phi}^-(k)) + \dot{\mathbf Z}(k)(\boldsymbol \phi(k)-\hat {\boldsymbol \phi}^-(k))\label{eqn:phiTaylor}
\end{eqnarray}
where the $N_r\times(N_r+N_t-1)$ Jacobian matrix with respect to $\boldsymbol \phi$ at time instance $k$, $\dot{\mathbf Z}(k)$, is defined as
\begin{eqnarray}
\dot{\mathbf Z}(k)&\triangleq&  \frac{\partial {\mathbf z}}{\partial \phi(k)}\Big |_{\hat {\boldsymbol \phi}^-(k)}.
\end{eqnarray}
The Jacobian matrix can be constructed with two matrices as
\begin{eqnarray}
\dot{\mathbf Z}(k) &=& [\dot{\mathbf Z}_1(k), \dot{\mathbf Z}_2(k)]
\end{eqnarray}
where the $N_r\times N_r$ matrix $\dot{\mathbf Z}_1(k)$ is defined as
\begin{eqnarray}
 \dot{\mathbf Z}_1(k) &\triangleq& \textrm{diag}\Big\{j z_1(\hat {\boldsymbol \phi}^-(k)),\dots,j z_{N_r}(\hat {\boldsymbol \phi}^-(k))\Big\}
\end{eqnarray}
and  the $N_r\times(N_t-1) $ matrix $\dot{\mathbf Z}_2(k)$ is given by   
\begin{align}
\label{eqn_dbl_x}
\dot{\mathbf Z}_2(k) \!\triangleq\!\left[\!\! \begin{array}{ccc} \!\!
  jh_{11}e^{j\hat {\phi}^{^-}_1(k)}\hat {\mathbf s}_1(k)e^{j\hat {\phi}^{^-}_{N_r+1}(k)}  \!&\!\! \dots \!\!&\! jh_{1(N_t-1)}e^{j\hat {\phi}^{^-}_{1}(k)}\hat {\mathbf s}_{N_t-1}(k)e^{j\hat {\phi}^{^-}_{N_r+N_t-1}(k)} \\
 \!\!jh_{21}e^{j\hat {\phi}^{^-}_2(k)}\hat {\mathbf s}_1(k)e^{j\hat {\phi}^{^-}_{N_r+1}(k)}\!&\!\! \dots \!\!&\! jh_{2(N_t-1)}e^{j\hat {\phi}^{^-}_{2}(k)}\hat {\mathbf s}_{N_t-1}(k)e^{j\hat {\phi}^{^-}_{N_r+N_t-1}(k)} \\
\!\!\vdots   \!&\!\! \ddots \!\!&\! \vdots  \\
   \!\! jh_{N_r1}e^{j\hat {\phi}^{^-}_{N_r}(k)}\hat {\mathbf s}_1(k)e^{j\hat {\phi}^{^-}_{N_r+1}(k)}  \!&\!\! \dots \ \!\!&\! jh_{N_r(N_t-1)}e^{j\hat {\phi}^{^-}_{N_r}(k)}\hat {\mathbf s}_{N_t-1}(k)e^{j\hat {\phi}^{^-}_{N_r+N_t-1}(k)}
\end{array} \!\!\right].
\end{align}

After the observation, the  a posteriori estimate of the state vector, $\hat {\boldsymbol \phi}(k)$,  and the error covariance matrix, $\hat {\mathbf M}(k)$, are given by
\begin{eqnarray}
\hat {\boldsymbol \phi}(k)&=&\hat {\boldsymbol \phi}^-(k) + \Re \{\mathbf K(k)(\mathbf y(k)-\mathbf z(\hat{\boldsymbol \phi}^-(k))) \}\\
\hat {\mathbf M}(k) &=& \Big(\mathbf I-\Re \{\mathbf K(k) \dot{\mathbf Z}(k)\}\Big)\hat {\mathbf M}^{^-}(k)\label{eqn:hatMpostMIMO},
\end{eqnarray}
where the $(N_r+N_t-1)\times N_r$ Kalman Gain matrix, $\mathbf K(k)$,  is given by
\begin{eqnarray}
 \mathbf K(k)&=& \hat {\mathbf M}^{^-}(k)\dot{\mathbf Z}(k)^H\Big(\mathbf C_w+\dot{\mathbf Z}(k) \hat {\mathbf M}^{^-}(k)\dot{\mathbf Z}(k)^H \Big)^{-1}.
\end{eqnarray}
Finally, the observation noise covariance matrix is given by
\begin{eqnarray}
\mathbf C_w&=&  \Big(\frac{\sigma^2_w}{2} + j\frac{\sigma^2_w}{2}\Big) \mathbf I.\label{eqn:hatCMIMO}
\end{eqnarray}

\begin{figure}[t]
\begin{center}
\includegraphics[width=12cm]{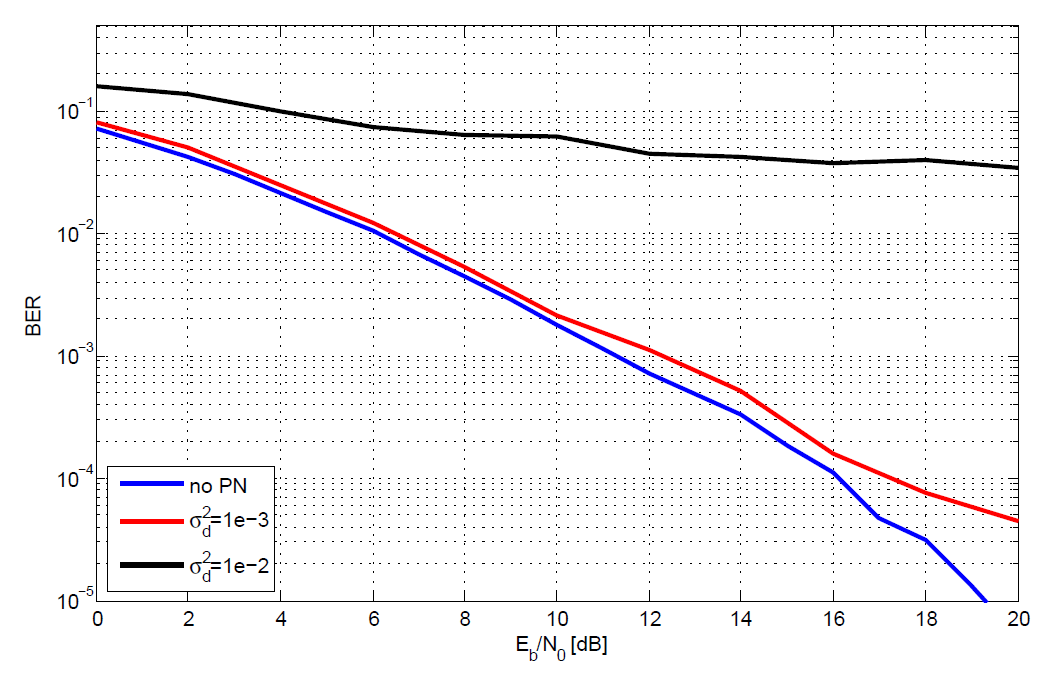}\center \caption{BER vs. $E_b/N_0$ for 2x2 uncoded MIMO system with BPSK modulation for various $\sigma_\Delta^2$ values.}\label{fig:fig2MIMO}
\end{center}
\end{figure}

\subsection{Simulation Results}
BER vs. $E_b/N_0$ for BPSK modulation and ML equalization for a 2x2 MIMO system where $L_f=1000$ is shown in Fig. \ref{fig:fig2MIMO}. It is seen that the EKF performs close to the synchronized system at low to medium  $E_b/N_0$ levels where $\sigma_\Delta^2=10^{^{-3}}$. On the other hand, the BER performance of the system degrades at high $E_b/N_0$ levels. Note that the number of parameters to be tracked by the EKF is larger than the number of observation equations. This results in an error floor, i.e. the performance of the system cannot be improved with increasing $E_b/N_0$. The error floor depends on the phase noise innovation variance, $\sigma_\Delta^2$. In Fig. \ref{fig:fig2MIMO} we observe that error floor tends to occur at lower signal-to-ratios for the system with $\sigma_\Delta^2=10^{^{-2}}$.

\newpage

\section{Coded MIMO}\label{sec:CodedMIMO}

MIMO systems can utilize the available spectrum more efficiently \cite{Telatar99},\cite{Foschini98}. A number of coding structures have been proposed in order to achieve high spectral efficiency, including multi-dimensional trellis-coded modulation \cite{Tarokh99}, space time block codes \cite{Alamouti}, MIMO multilevel coding \cite{Lampe}.   BICM is also one of the popular scheme that enables communication systems to fully exploit the spectrum efficiently \cite{Duman,Boutros,Sellathurai}.

 Code-aided synchronization based on the EM framework for joint channel estimation, frequency and time synchronization for a BICM-MIMO system is proposed in \cite{HenkSimoens}. However, in \cite{HenkSimoens}, the synchronization parameters are assumed to be constant and deterministic over the length of a block, which is not a valid assumption for time varying phase noise. In the following, we show analytically how an EM-based algorithm  with a soft decision-directed EKFS can be used to estimate and compensate time varying phase noise.

 First, we recall the signal model for uncoded MIMO systems. The received signal at time instance $k$ is given by,
 \begin{eqnarray}
 \mathbf y(k) &=& \boldsymbol \Gamma^{[r]}(k) \mathbf H \boldsymbol \Gamma^{[t]}(k) \mathbf s(k) + \mathbf w(k)\nonumber\\
 &=& \mathbf X(k) {\mathbf s}(k) + \mathbf w(k)\nonumber
\end{eqnarray}
and the new state vector of length $N_t+N_r-1$ can be written as
\begin{eqnarray}
\boldsymbol{\phi}(k)&=&\boldsymbol{\phi}(k-1)+\boldsymbol{\Delta}(k)\nonumber.
\end{eqnarray}

\subsection{The EM algorithm\label{sec:EMalg}}
The EM algorithm iteratively solves an estimation problem where the observation, $\left[\mathbf Y\right]_{N_r \times L_f} \triangleq [\mathbf y(1) , \mathbf y(2) , \dots ,\mathbf y(L_f)]$, depends not only on the parameters to be estimated, $\left[\boldsymbol \Theta\right]_{(N_r+N_t) \times L_f} \triangleq [\boldsymbol \theta(1) , \boldsymbol \theta(2) , \dots ,\boldsymbol \theta(L_f)]$, but also on some unknown parameters, $\left[\mathbf S\right]_{N_t \times L_f} \triangleq [\mathbf s(1) , \mathbf s(2) , \dots ,\mathbf s(L_f)]$ \cite{Moon}.

Given the observation sequence $\mathbf Y$, the MAP estimate of $\hat{\boldsymbol \Theta}$ is given by \cite{SMKay}\vspace{-3pt}
\begin{align}
\hat{ \boldsymbol \Theta}= \textrm{arg}\, \max_{ { \boldsymbol \Theta}} \ln\ p(\mathbf Y| {\boldsymbol \Theta})+\ln p({ \boldsymbol \Theta})
\end{align}
where $p(\mathbf Y| {\boldsymbol \Theta})=\displaystyle\sum_{\mathbf S} p(\mathbf S) p(\mathbf Y|\mathbf S, {\boldsymbol \Theta})$.

The EM algorithm consists of the \emph{expectation step }(E-step) and \emph{maximization step} (M-step) \cite{Herzet1}. For the $i$th EM iteration, the E-step and M-step equations are given by
\begin{eqnarray}
\label{eq:E-stepMIMO}
\mathrm{Q}\Big( {\boldsymbol \Theta}| \hat{\boldsymbol \Theta}^{(i-1)}\Big) &\triangleq&
 \mathbb{E}_{\mathbf S|\mathbf Y,\hat{\boldsymbol \Theta}^{(i-1)}}\Big\{\ln p(\mathbf Y|\mathbf S, {\boldsymbol \Theta})\Big\}+\ln p(\boldsymbol \Theta), \\
 \label{eq:M-stepMIMO}
 \hat{\boldsymbol \Theta}^{(i)}&=&\arg\max_{ {\boldsymbol \Theta}}\Big\{\mathrm{Q}\Big( {\boldsymbol \Theta}| \hat{\boldsymbol \Theta}^{(i-1)}\Big)\Big\},
\end{eqnarray}
respectively.
The EM algorithm converges to the MAP solution if the initial estimates of the parameters of interest,
$\hat{\mathbf b}^{(0)}$, are sufficiently close to the true values of the parameters. Otherwise, the EM algorithm may converge to a saddle point or a local maximum. To ensure the convergence, pilot symbols are inserted into the data stream every $p_r$ time instances.

In the following subsections we derive the E-step and M-step of the EM algorithm for coded MIMO systems.


\subsection{E-Step \label{sec:Estep}}
Let us define $\mathbf X(k) \triangleq   \boldsymbol \Gamma^{[r]}(k) \mathbf H \boldsymbol \Gamma^{[t]}(k)$. Consequently, the vector of received signals at the receive antennas at time $k$, $\mathbf y(k)$ can be rewritten as
\begin{align}\label{eq:received_signal2}
 \mathbf y(k) =&  \mathbf X(k) \mathbf s(k) + \mathbf w(k).
\end{align}
According to \eqref{eq:received_signal2}, the LLF of the received signal matrix, $\mathbf{Y}$, given the transmitted data, $\mathbf S$, and the phase noise process, $ {\boldsymbol \Theta}$, is given by
\begin{align}
\ln p(\mathbf Y|\mathbf S, {\boldsymbol \Theta})\varpropto& -\sum_{k=1}^{L_f} \parallel \mathbf y(k)- \mathbf X(k) \mathbf s(k) \parallel\notag\\
\varpropto& 2\Re \sum_{k=1}^{L_f} \textrm{tr}\Big(\mathbf y(k) \mathbf s^H(k) \mathbf X^H(k) \Big) \nonumber\\
&\hspace{1cm}- \sum_{k=1}^{L_f} \textrm{tr}\Big(\mathbf X(k) \mathbf s(k) \mathbf s^H(k) \mathbf X^H(k) \Big).\label{eqn:channelLF}
\end{align}
Using \eqref{eqn:channelLF} in \eqref{eq:E-stepMIMO} the E-step of the EM algorithm can be determined as
\begin{align}
\mathrm{Q}\Big( {\boldsymbol \Theta}|\hat{\boldsymbol \Theta}^{(i-1)}\Big)
=&2\Re \sum_{k=1}^{L_f} \textrm{tr}\Big(\mathbf y(k) \boldsymbol \alpha^H(k) \mathbf X^H(k) \Big) \nonumber \\
&- \sum_{k=1}^{L_f} \textrm{tr}\Big(\mathbf X(k)  \mathbf B(k) \mathbf X^H(k) \Big)
+\ln p(\mathbf {\boldsymbol \Theta}), \label{eqn:Estepfinal}
\end{align}
where
\begin{itemize}
\item $\boldsymbol \alpha(k)\hspace{-.05cm}\triangleq \hspace{-.05cm}\mathbb{E}_{\mathbf s}\{\mathbf s(k)\}= \displaystyle\sum_{\mathbf {a}_n \in  \Omega^{N_t}}\mathbf {a}_n p(\mathbf {s}(k)=\mathbf {a}_n|\mathbf Y,\hat{\boldsymbol \Theta}^{(i-1)})$ and
\item $\mathbf B(k)\hspace{-.05cm}\triangleq\hspace{-.05cm} \mathbb{E}_{\mathbf s}\{\mathbf s(k) \mathbf s^H(k)\}
= \displaystyle\hspace{-.4cm}\sum_{\mathbf {a}_n \in  \Omega^{N_t}}\hspace{-.3cm}\mathbf {a}_n \mathbf {a}^H_n p(\mathbf {s}(k)\hspace{-.05cm}=\hspace{-.05cm}\mathbf {a}_n|\mathbf Y,\hat{\boldsymbol \Theta}^{(i-1)})$.
\end{itemize}
In \eqref{eqn:Estepfinal}, $\boldsymbol \alpha(k)$ denotes the marginal posterior mean of the coded symbol vector at time instance $k$, $p(\mathbf {s}(k)=\mathbf {a}_n|\mathbf Y,\hat{\boldsymbol \Theta}^{(i-1)})$ denotes the APPs of the coded symbol vector given $\mathbf Y$ and $\hat{\boldsymbol \Theta}^{(i-1)}$,
and $\mathbf B(k)$ is an ($N_t\times N_t$) matrix. As a result, the E-step reduces to the computation of the APPs.  Note that the APPs can be computed via an iterative MAP decoder which is outlined in Section \ref{sec:Apps}.

\subsection{M-Step\label{sec:Mstep}}

As shown in \eqref{eq:M-stepMIMO}, the M-step maximizes the E-step with respect to the parameter of interest, $\boldsymbol\Theta$. In the following, it is shown that a low complexity soft decision-directed EKFS can be used to carry out the M-step of the EM algorithm.

Note that a Kalman filter is equivalent to a MAP estimator \cite{SMKay}. As a result, the Kalman filter estimates of $\boldsymbol\Theta$, $\hat{\boldsymbol\Theta}$ can be written as\vspace{-4pt}
\begin{align}
\hat{ \boldsymbol{\Theta}}
=& \textrm{arg}, \max_{ { \boldsymbol{\Theta}}} \{\ln p(\mathbf Y| {\boldsymbol{\Theta}}, \mathbf S= \mathbf A) + \ln p(\boldsymbol{\Theta})\}\label{eqn:MAPln2}
\end{align}
where  $\mathbf A\triangleq  [\boldsymbol \alpha(1) , \boldsymbol \alpha(2) , \dots ,\boldsymbol \alpha(L_f)]$ is an $N_t\times L_f$ matrix.

By setting $\mathbf{S}= \mathbf A$, (\ref{eqn:channelLF}) can be rewritten as
\begin{align}
\ln p(\mathbf Y|\mathbf S=\mathbf A, {\boldsymbol \Theta})
\varpropto& 2\Re \sum_{k=1}^{L_f} \textrm{tr}\Big(\mathbf y(k) \boldsymbol \alpha^H(k) \mathbf X^H(k) \Big) \nonumber \\
&- \sum_{k=1}^{L_f} \textrm{tr}\Big(\mathbf X(k) \boldsymbol \alpha(k) \boldsymbol \alpha^H(k) \mathbf X^H(k) \Big)\nonumber\\
&+\ln p(\boldsymbol{\Theta})\label{eqn:channelLF2}.
\end{align}
Note that (\ref{eqn:channelLF2}) and the E-step equation in (\ref{eqn:Estepfinal}) are equivalent to one another except for the term $\displaystyle\sum_{k=1}^{L_f} \textrm{tr}\Big(\mathbf X(k) \boldsymbol \alpha(k) \boldsymbol \alpha^H(k) \mathbf X^H(k) \Big)$. However, when the soft decisions reach their true values, we can assume that $\boldsymbol \alpha(k)\boldsymbol \alpha^H(k)\approx \mathbf B(k)$. Thus, using \eqref{eqn:MAPln2} and \eqref{eqn:channelLF2}, we can conclude that a Kalman filter can be applied to carry out the M-step of the EM algorithm.

Note that the \emph{observation} equation for the Kalman filter is given by \eqref{eq:received_signal2} while based on \eqref{eq:pn_model}, the\emph{ state} equation at time $k$ is given by\vspace{-4pt}
\begin{align}\label{eqn:state3}
  \boldsymbol{\theta}(k)=&\boldsymbol{\theta}(k-1)+\boldsymbol{\Delta}(k),
\end{align}
where $\boldsymbol{\Delta}(k)\triangleq\left[ \Delta_1^{[r]}(k), \dots, \Delta_{N_r}^{[r]}(k), \Delta_1^{[t]}(k), \dots, \Delta_{N_t}^{[t]}(k)\right]^T$. The transmitted symbols $\mathbf s(k)$, can be replaced by their a posteriori means, i.e., soft decisions $\boldsymbol \alpha(k)$ computed at the E-step using the iterative MAP detector in Section \ref{sec:Apps}. Subsequently, the \emph{observation} equation in \eqref{eq:received_signal2} can be rewritten as
\begin{eqnarray}\label{eqn:received_approxMc}
 \mathbf y(k) \approx& \mathbf X(k) \boldsymbol \alpha(k) + \mathbf w(k).
\end{eqnarray}
Since the \emph{observations} are a nonlinear function of the parameters of interest, $\boldsymbol{\theta}_k$, an \emph{extended} Kalman filter-smoother needs to be used instead to carry out the M-step of the EM   algorithm \cite{SMKay}.

\subsection{The Extended Kalman Filter-Smoother\label{sec:EKFS}}

Note that (\ref{eq:obsMIMOunc2}) and (\ref{eqn:received_approxMc}) are similar. We can rewrite the observation equation for the coded MIMO system as
\begin{eqnarray}
y(k)&=& \mathbf z_c(\boldsymbol{\phi}(k))+w(k).
\end{eqnarray}

The EKFS first filters the received signal and provides phase noise estimates with a forward recursion over the frame.
The filtering equations compute the posteriori estimate of the state vector, $\hat {\boldsymbol \phi}^+(k)$,  and the error covariance matrix, $\hat {\mathbf M}^{^+}(k), k=2,3,\dots,L_f$ in a recursive fashion according to (\ref{eqn:hatPhiMIMO}-\ref{eqn:hatMpostMIMO}) where $\mathbf z(\boldsymbol{\phi}(k))$ is replaced by $\mathbf z_c(\boldsymbol{\phi}(k))$. In other words, instead of the hard decision vector provided by the ML detector in the  uncoded MIMO system, $\hat {\mathbf s}(k)$, the soft decision vector computed by the iterative detector, $\alpha(k)$, is provided to the EKFS for the coded MIMO system. The EKFS is initialized with the state estimate $\hat {\boldsymbol \phi}(0)=0$ and the error covariance estimate $\hat {\mathbf M}(0)=\sigma^2_\Delta\mathbf I$.

\begin{figure}[t]
\begin{center}
\includegraphics[width=14cm]{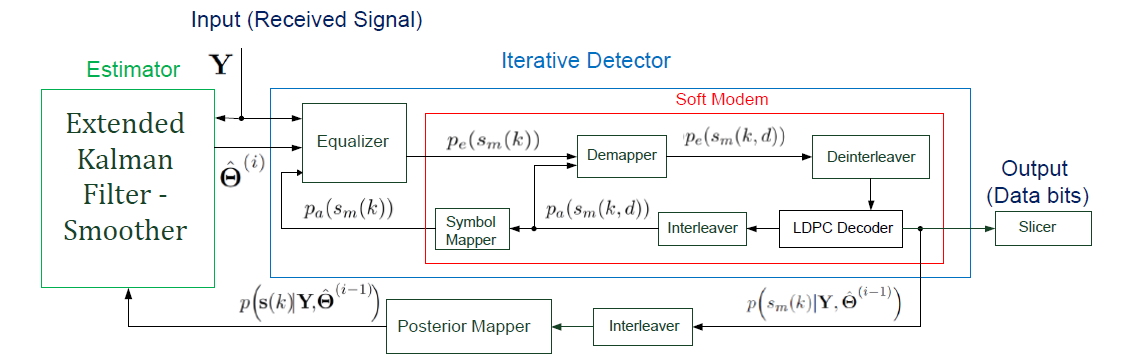}
\caption{Block diagram of the receiver structure.}\label{fig:fig3}
\end{center}
\vspace{-.3cm}
\end{figure}

%
%

%
 After filtering is completed for the whole block, a backward recursion is performed to smooth the estimates. Finally, the smoothed estimates of the phase noise parameters are carried out to be fed to the detector.
The smoothed estimate of the phase noise vector, $\hat {\boldsymbol \phi}(k)$, and the error covariance matrix $\hat {\mathbf M}(k)$ are given by
\begin{eqnarray}
\!\!\hat {\boldsymbol \phi}(k)\!\!\!&=&\!\!\!\hat {\boldsymbol \phi}^+(k) \!+ \hat {\mathbf M}^{^+}(k) (\hat {\mathbf M}^{^-}(k))^{-1}\Big(\hat {\boldsymbol \phi}(k+1)\!-\!\hat {\boldsymbol \phi}^+(k)\Big) \\
\!\!\hat {\mathbf M}(k)\!\!\!&=&\!\!\!\hat {\mathbf M}^{^+}(k)\!+\hat {\mathbf M}^{^+}(k) (\hat {\mathbf M}^{^-}(k))^{-1}\Big(\hat {\mathbf M}(k+1)\!-\hat {\mathbf M}^{^+}(k)\Big)\nonumber\\
  \!\!&&\!\! \times  \Big(\hat {\mathbf M}^{^+}(k) (\hat {\mathbf M}^{^-}(k))^{-1}\Big)^T.
\end{eqnarray}
After the backward recursion is completed the block of phase noise estimates $\hat {\boldsymbol \phi}(k),k=1,2,\dots,L_f$, is fed to the iterative detector for the next EM algorithm iteration.

\subsection{Iterative Detector\label{sec:Apps}}
It is shown in Sec. \ref{sec:Mstep} that soft decisions, i.e., the marginal posterior probabilities of the coded symbol vectors $\mathbf A$ are required for the EKFS. The computation of the true posterior probabilities has a complexity that increases exponentially with the frame length $L_f$. Therefore, a near optimal iterative detector, operating according to the turbo principle  \cite{Duman},\cite{Boutros},\cite{HenkSimoens} is used to obtain the marginal a posteriori bit probabilities given the phase noise estimates and the channel gain matrix $\mathbf H$. Then, a soft modulator maps the a posteriori bit probabilities to symbol probabilities and constructs the soft decisions. The block diagram of the proposed EM-based receiver structure, including both the EKFS and the iterative detector, is shown in Fig. \ref{fig:fig3}.

The iterative detector first computes $M^{N_t}$ conditional likelihoods of the symbol vectors given the phase noise estimates  as 
\begin{eqnarray}
\!\!p\Big(\!\mathbf Y|\mathbf s(k)\!,\!\hat{\boldsymbol \Theta}^{(i-1)}\!\Big)\!\!&\!\!=\!\!&\!\!p\Big(\!\mathbf y(k)|\mathbf s(k),\hat{\boldsymbol \Theta}^{(i-1)}\Big)\nonumber\\
\!\!&\!\!=\!\!&\!\!C'\!\exp\! \Big(\!-\!\frac{1}{2\sigma^2_w}|\mathbf y(k)\!-\!\mathbf X(k)\mathbf s(k)|^2 \Big).
\end{eqnarray}
Note that the conditional likelihoods are computed by the equalizer from  the received signal before the execution of the detector iterations. Iterative detection is then performed with 3 nested iterations, as seen in Fig. \ref{fig:fig3}.

The iterative part of the detector operates according to the turbo principle.
The conditional \emph{a posteriori} probability of the transmitted symbol at the transmit antenna $m$, at time instance $k$, $s_m(k)$, is factored as the product of an \emph{a priori} probability (with subscript $a$), and an \emph{extrinsic} probability (with subscript $e$) such as
\begin{eqnarray}
p\Big(s_m(k)|\mathbf Y, \hat{\boldsymbol \Theta}^{(i-1)}\Big)&=&C^{(1)}p_a(s_m(k))p_e(s_m(k))\label{eqn:symbPostFact}
\end{eqnarray}
where $C^{(1)}$ is a normalization constant.
Note that since an interleaver at the transmitter side is used, the transmitted symbols on each antenna and the transmitted bits within each constellation symbol are independent.
The extrinsic symbol probabilities in (\ref{eqn:symbPostFact}) are computed by the equalizer in Fig. \ref{fig:fig3}. The extrinsic symbol probability given the estimated phase noise parameters and the received signal  $p_e(s_m(k)=a_n)$ for $a_n\in \Omega, n=1,2,\dots,M$, is given by
\begin{eqnarray}
p_e(s_m(k)=a_n)&=&p\Big(\mathbf Y|s_m(k)=a_n,\hat{\boldsymbol \Theta}^{(i-1)}\Big) \nonumber\\
     &=& \sum_{\mathbf s(k):s_m(k)=a_n} \Big\{p\Big(\!\mathbf Y|\mathbf s(k),\hat{\boldsymbol \Theta}^{(i-1)}\Big)\nonumber\\
      &&\qquad\times\prod_{m'\neq m} p_a(s_{m'}(k))\Big\}.
\end{eqnarray}
The a priori symbol probabilities $p_a(s_m(k,d))$ are computed by the soft modem and fed to the equalizer after decoding is performed. Therefore, the first iteration is between the equalizer and the soft modem and there are $L_{eq-sm}$ iterations.
The soft modem consists of the demapper, the interleaver, the iterative MAP decoder, the deinterleaver and the mapper as it is shown in Fig. \ref{fig:fig3}. The demapper takes the extrinsic \emph{symbol} probabilities from the equalizer and computes the extrinsic \emph{bit} probabilities. The bit posterior probabilities of the $d$th bit of the bit sequence mapped to the symbol $s_m(k)$, denoted as $s_m(k,d)$, is factored similar to (\ref{eqn:symbPostFact}) as
\begin{eqnarray}
p\Big(\!s_m(k,d)|\mathbf Y\!,\hat{\boldsymbol \Theta}^{(i-1)}\!\Big)\!=\!C^{(2)}p_a(s_m(k,d))p_e(s_m(k,d)).
\end{eqnarray}
where $C^{(2)}$ is a normalization constant.
Then, the extrinsic bit probability for $\beta \in \{0,1\}$ is computed by the demapper as
\begin{eqnarray}
p_e(s_m(k,d)=\beta)&=&p\Big(\mathbf Y|s_m(k,d)=\beta,\hat{\boldsymbol \Theta}^{(i-1)}\Big) \nonumber\\
     &=& \sum_{a_n \in \Omega:a_n(d)=\beta} \Big\{p_e(s_m(k)=a_n)  \nonumber\\
     && \qquad\prod_{d'\neq d} p_a(s_m(k,d'))\Big\}.
\end{eqnarray}
The demapper also needs the a priori \emph{bit} probabilities computed by the MAP decoder. The second iteration of the 3 nested iterations of the detector is performed  between the demapper and the decoder $L_{dm-dc}$ times. After the demapper computes all the extrinsic bit probabilities, they are deinterleaved and provided to the MAP decoder by the deinterleaver. The decoder takes the deinterleaved extrinsic bit probabilities and computes both the deinterleaved \emph{posterior} and the deinterlaved \emph{a priori} bit probabilities exploiting code properties with $L_{dec}$ iterations. The posterior bit probabilities $p\Big(s_m(k,d)|\mathbf Y,\hat{\boldsymbol \Theta}^{(i-1)}\Big)$ are sent out of the detector, interleaved and used to construct soft decisions for the EKFS, described in Section \ref{sec:EKFS}. The posterior bit probabilities are also used for hard decision after the algorithm terminates.
The a priori bit probabilities $p_a(s_{n}(k,d))$ are sent to the interleaver which constructs the a priori bit probabilities that are used as a priori information by the demapper. Finally, the mapper converts a priori bit probabilities to a priori symbol probabilities which are used as a priori information by the equalizer according to
\begin{eqnarray}
p_a(s_m(k,d))&=&\prod_{d} p_a(s_{n}(k,d)).
\end{eqnarray}
\begin{figure}[t]
\begin{center}
\includegraphics[width=9cm]{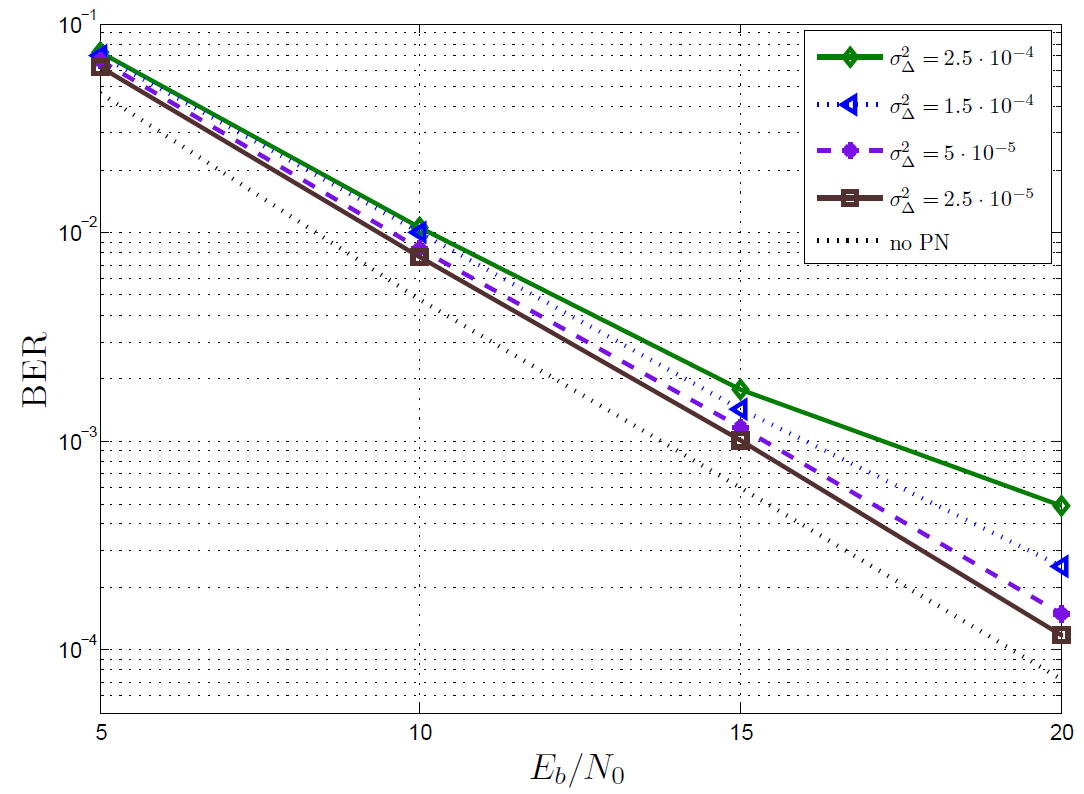}\center \caption{BER performance of the EM-based algorithm with $DA(14)$ estimator.}\label{fig:BERpr14}
\end{center}
\end{figure}

Aforementioned, the iterative detector provides the EKFS with the marginal posterior probabilities of the coded symbol vectors, $p\Big(\!\mathbf s(k)\!|\mathbf Y\!,\! \hat{\boldsymbol \Theta}^{(i-1)}\!\Big)$. The E-step is finished by the computation of these probabilities  as
\begin{eqnarray}
\!\!\!p\Big(\!\mathbf s(k)\!|\mathbf Y\!,\! \hat{\boldsymbol \Theta}^{(i-1)}\!\Big)\!\!\!&\!\!=\!\!&\!\!\!C^{(3)}p_a(\mathbf s(k))p_e(\mathbf s(k)) \\
\!\!\!&\!\!=\!\!&\!\!\!C^{(4)}p\Big(\!\mathbf Y\!|\mathbf s(k)\!,\!\hat{\boldsymbol \Theta}^{(i-1)}\!\Big)\!\prod_{m,d} p_a(s_m(k,d))
\end{eqnarray}
where $C^{(3)}$ and $C^{(4)}$ are  normalization constants.
Then, these posterior probabilities are used to construct the soft decision, $\boldsymbol \alpha(k)$.

The a priori symbol probabilities $p_a(s_m(k))$ and bit probabilities $p_a(s_m(k,d))$ are initialized with a uniform distribution at the first EM algorithm iteration. A sufficient number of iterations  inside of the detector is required  for convergence if the detector is reinitialized with uniform probabilities at each EM iteration. Instead, the detector can be initialized with the a priori probabilities obtained at the previous EM iteration. In addition, only 1 iteration is allowed inside of the iterative detector for each nested iteration, i.e., $L_{eq-dm}=L_{dm-dc}=L_{dec}=1$ as suggested in \cite{HenkSimoens}. This alternative approach requires more EM iterations but less detector iterations to converge. In this way, the computational complexity of both the detector and the receiver can be reduced significantly.

\subsection{Simulation Results\label{sec:SimRes}}

\begin{figure}[t]
\begin{center}
\includegraphics[width=9cm]{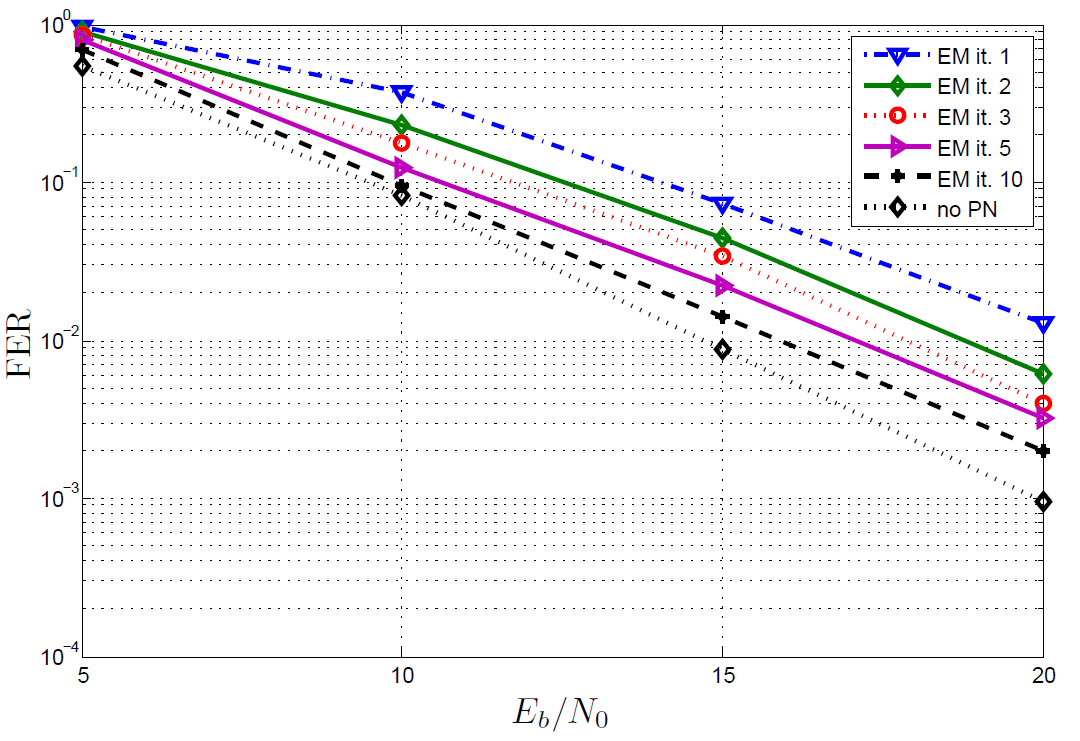}\center \caption{FER performance of the EM-based algorithm at several iterations where $\sigma^2_\Delta=5\cdot10^{-5}$, and DA initial estimation with $p_r=14$.}\label{fig:fer2EMits}
\end{center}
\end{figure}

At the transmitter, data bits are first encoded by a rate $R=7/8$ regular LDPC encoder  from the NASA Goddard technical standard \cite{Goddard}. It is a regular code with variable node degree 4 and check node degree 32.
  The number of data bits in each frame, $L_b$  is equal to 7154. Then, encoded bits are modulated onto 16-QAM symbols. Therefore, there are $L_f=L_b/N_tR\log_2(M)=1022$ symbol vectors in each frame.
 Performance will be measured as a function of $E_b/N_0$, where $E_b$ denotes the transmitted energy per information bit and $N_0$ is the power spectral density of the AWGN, i.e, $\sigma^2_w=N_0$.

 The convergence of the EM based estimator is severely dependent on the initialization of the estimation parameters. First, the EM based algorithm is initialized by a data aided (DA) estimator to provide the initial phase noise estimates. Pilots are inserted into the data stream every $p_r$ time instance.  At the first EM algorithm iteration, the EKFS makes use of the pilots and estimates phase noise values at each $p_r$ time instance. Afterwards a linear interpolation is performed between two consecutive phase noise estimates. As a result, initial phase noise estimates are obtained and sent to the iterative detector to initialize the EM-based algorithm. DA$(p_r)$ denotes the DA estimator with pilot rate $p_r$.

\begin{figure}[t]
\begin{center}
\includegraphics[width=9cm]{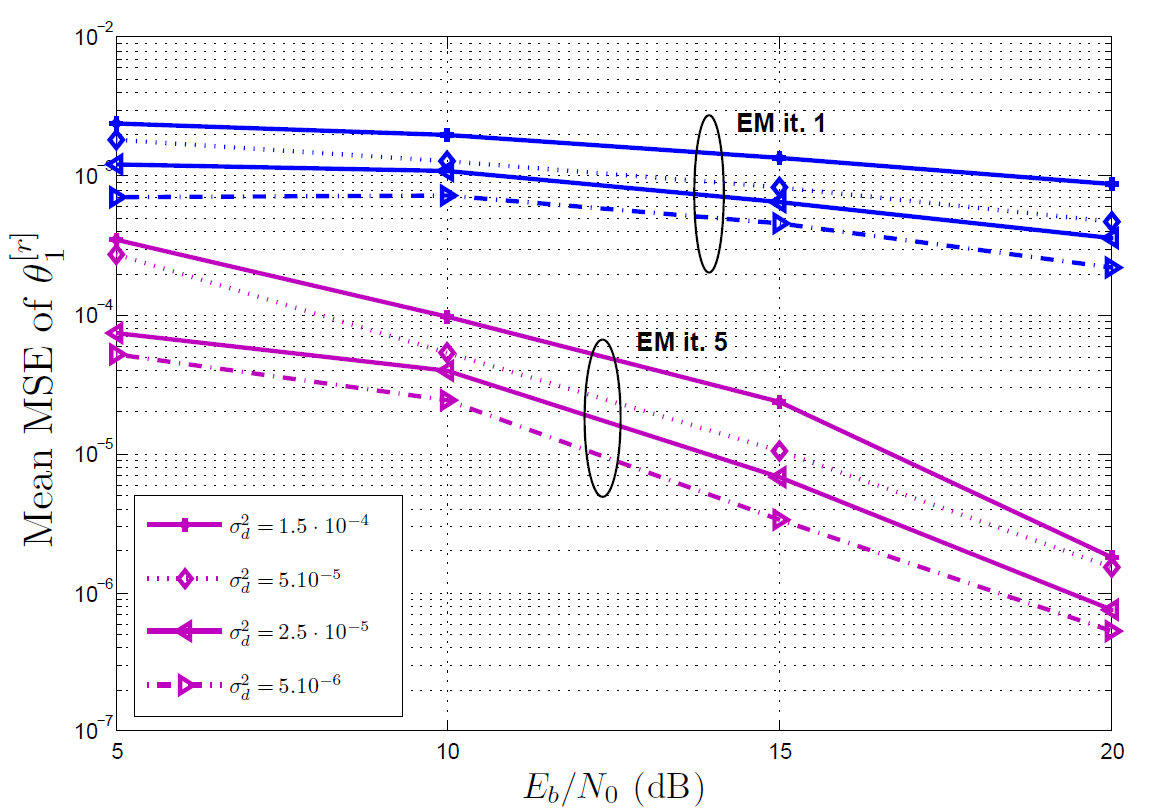}\center \caption{MSE performance of the EM-based algorithm, and DA initial estimation with $p_r=14$ for several phase noise processes and EM algorithm iterations.}\label{fig:mseEMits}
\end{center}
\end{figure}

In Fig. \ref{fig:BERpr14}, the effect of the phase noise on the BER performance corresponding to the EM-based algorithm with DA$(14)$ estimator for different phase noise innovation variance levels, $\sigma^2_\Delta$,  after the 3rd EM algorithm iteration is shown.  As compared to perfect synchronization, no phase noise scenario, the proposed EM-based algorithm gives rise to a BER degradation of about 1dB in the presence of the slowly time varying phase noise, i.e., $\sigma^2_\Delta=2.5\cdot10^{-5}$. The BER degradation amounts to about 2dB when $\sigma^2_\Delta=1.5\cdot10^{-4}$. It is possible to track a stronger phase noise process with innovation variance $\sigma^2_\Delta=2.5\cdot10^{-4}$ in expense of 4dB of $E_b/N_0$ comparing to perfectly synchronized system. Note that the performance of the system tends to stay constant as $E_b/N_0$ increases, yielding an error floor. The main reason is that the prediction error of the EKFS is determined by the phase noise innovation variance. As $\sigma^2_\Delta$ increases, the error floor is observed at higher error rates.
\begin{figure}[t]
\begin{center}
\includegraphics[width=9cm]{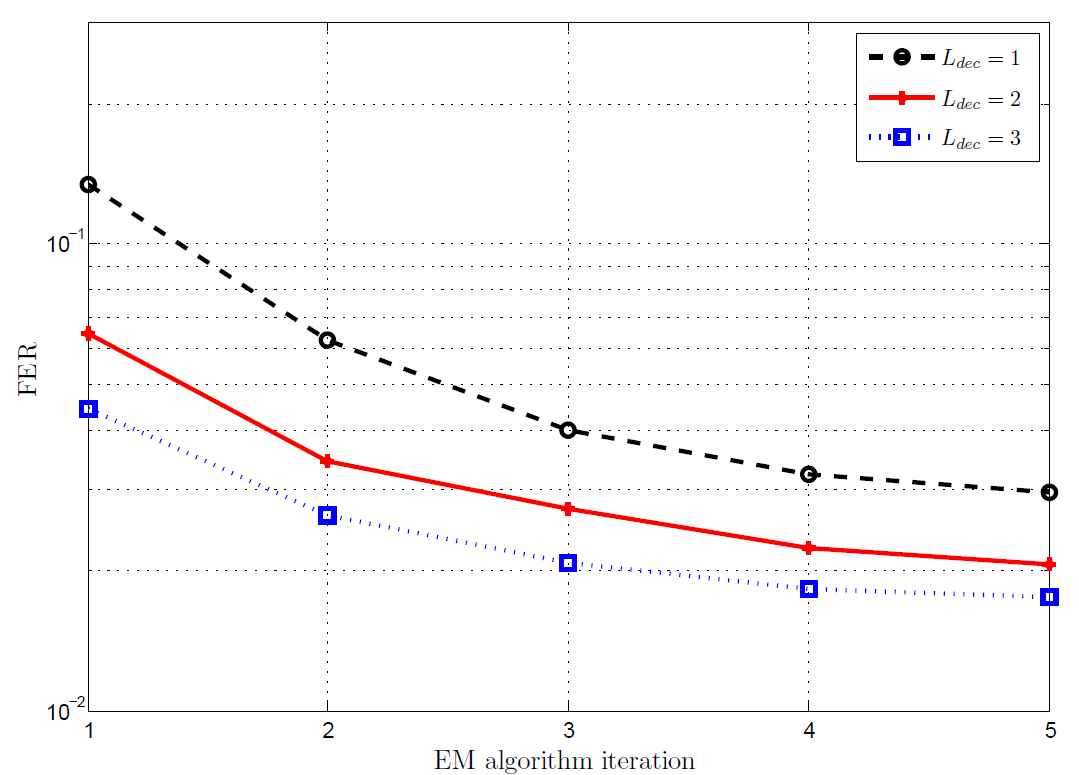}\center \caption{FER performance of the EM-based algorithm, and DA initial estimation with $p_r=14$ for several number of decoder iterations, $\sigma^2_\Delta=5\cdot10^{-4}$.}\label{fig:ferLdec}
\vspace{-.5cm}
\end{center}
\end{figure}

Code-aided synchronization techniques offer to improve the overall accuracy and performance of the system. First, we investigate the performance of the EM-based algorithm at each EM algorithm iteration. The EM-based algorithm does not converge to the global solution at some of the erroneous frames resulting in an arbitrary large number of erroneous bits. For this reason the BER performance of the system may not increase at each EM algorithm iteration. Therefore,  the FER performance of the system is investigated. Fig. \ref{fig:fer2EMits}
shows the FER performance at several EM algorithm iterations for $\sigma^2_\Delta=5\cdot10^{-5}$. We observe that the FER performance of the system improves at each EM algorithm iteration. In order to operate the system at $2\cdot10^{-3}$ FER at the 10th iteration, additional $E_b/N_0$ of about 2dB is required. Secondly, we investigate the estimation accuracy of the EM-based algorithm. The MSE of the phase noise estimates at the first receive antenna is averaged over the successfully decoded frames. Fig. \ref{fig:mseEMits} shows the MSE results of different levels of phase noise innovation variance at both the 1st and the 5th EM algorithm iteration. It is seen that the proposed EM-based algorithm not only increases the overall performance of the system but also yields better estimates at each EM algorithm iteration.

We also investigate the performance of the proposed EM-based algorithm for the system that is affected by a severe phase noise process such that $\sigma^2_\Delta=5\cdot10^{-4}$. Fig. \ref{fig:ferLdec} shows the effect of the number of decoder iterations on the FER performance of the system for a fixed $E_b/N_0=20dB$. We observe that the performance of the system does not improve significantly after a few EM algorithm iterations. In contrast, the FER performance of the system can be slightly improved with more iterations inside of the decoder. In order to achieve lower error rates in the presence of strong phase noise a stronger channel encoder can be used, i.e, the rate of the LDPC code, $R$, needs to be decreased. Note that the spectral efficiency of the system also reduces with decreasing $R$ yielding low throughput.

\begin{figure}[t]
\begin{center}
\includegraphics[width=9cm]{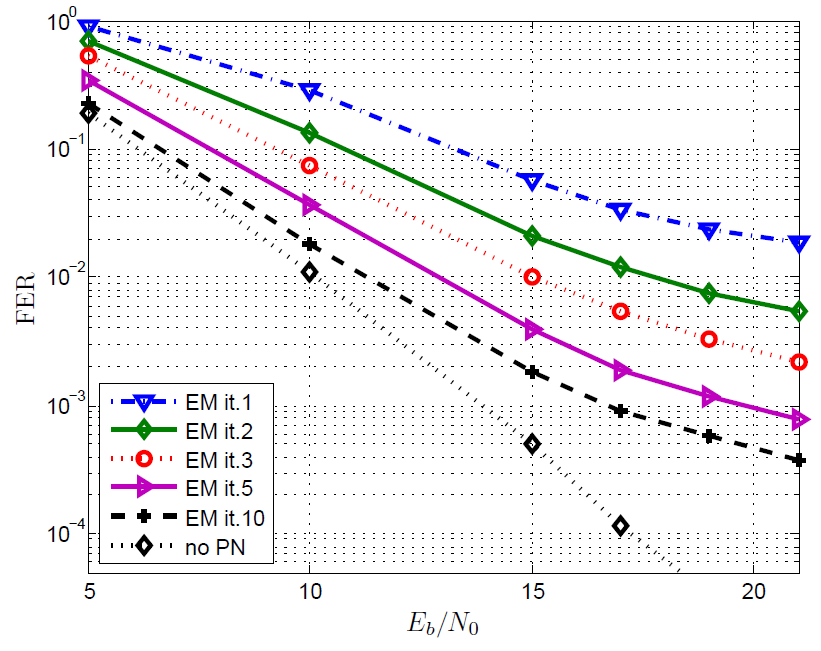}\center \caption{FER performance of the EM-based algorithm at several iterations where $\sigma^2_\Delta=5\cdot10^{-4}$, DA initial estimation with $p_r=14$, and $R=1/2$ rate code.}\label{fig:fer5e4rate12}
\end{center}
\end{figure}

Fig \ref{fig:fer5e4rate12} shows the FER performance of the EM-based algorithm for $\sigma^2_\Delta=5\cdot10^{-4}$ where data bits are encoded by a rate $R=1/2$
LDPC encoder \cite{MacKay}.  We observe that the FER performance can be improved significantly by increasing the number of EM iteration. It is possible to achieve  $5\cdot10^{-3}$ FER in expense of 4dB of $E_b/N_0$ comparing to no phase noise scenario. However, the error floor still occurs.
It is concluded that the number of the decoder iterations of the EM-based algorithm, $L_{dec}$, should be tuned to a sufficiently large number to be able to track time varying phase noise.


\chapter{Conclusion and Future Work}\label{chap:concFW}
\section{Conclusion}
In this thesis, phase noise estimation in uncoded/coded SISO and MIMO system is analyzed.

In Chapter \ref{chap:SISO-model}, phase noise tracking is performed by a hard decision directed EKF for the uncoded SISO system. Numerical results shows that the EKF is able to track slowly time varying phase noise processes. Unsurprisingly, the BER degradation reaches large values when the phase noise innovation variance is high. Additionally, in  Chapter \ref{chap:SISO-model}, the problem of joint phase noise estimation and detection in a coded SISO system with LDPC codes is discussed. The EM-based algorithm proposed in \cite{Shehata} is modified and analytically derived. Two estimators, a soft decision directed KS-MLA and a soft decision directed EKS proposed in \cite{Shehata} are applied to carry out the maximization step of the EM-based algorithm. In \cite{Shehata}, the KS-MLA is claimed to have superior performance than the EKS. However, numerical results in Chapter \ref{chap:SISO-model} indicates that when phase noise over the frame reaches very large values, i.e., in the case of large block length and/or phase noise innovation variance, the performance of the KS-MLA degrades significantly and the EKS performs better than the KS-MLA. The KS-MLA removes the data dependency by multiplying the observed signal with the soft decision symbol. As a result, the performance of the KS-MLA degrades faster than the EKS when the soft decisions are less reliable. For instance, for the same phase noise innovation variance, the EKFS performs better than the KS-MLA when the constellation density increases. A trick to increase the algorithm speed is also discussed and shown to decrease the overall complexity required for convergence.

In Chapter \ref{chap:MIMO-model}, a low complexity hard decision directed EKF is derived and applied to an uncoded MIMO system. Simulation results show that the EKF performs close to the synchronized system in the case of slowly time varying phase noise process.
 The problem of joint estimation of the time varying phase noise and data detection for a LOS-MIMO system using bit interleaved coded modulation and LDPC codes is also discussed. An iterative EM-based receiver to perform code-aided synchronization is proposed. A new low complexity soft-decision directed EKFS is derived and embedded into the EM-based algorithm. The performance of the system is  investigated in terms of the BER and the FER. Estimation accuracy of the phase noise parameters is presented with average MSE curves. Computational complexity of the system is also discussed. Computer simulations show that the number of bit errors does not decrease at each EM algorithm iteration if the algorithm fails to converge. Instead, the FER performance is shown to decrease at each EM iteration.
Simulation results demonstrate that the proposed EM-based algorithm estimates and compensates the time varying phase noise with a small degradation of the performance for a wide range of phase noise innovation variances. However, an error floor occurs at high signal-to-noise ratio levels. To reduce the error floor and to track the phase noise process with large innovation variance decoding performance can be increased by setting the number of decoder iterations to larger values. However, the achieved performance gain is not significant comparing to the introduced complexity. On the other hand, coding rate can be reduced further to protect data bits, yielding better soft decisions. The FER performance can be significantly improved by utilizing low rate LDPC codes at the expense of a decrease in throughput. As a result, the EKFS can be applied to a coded LOS-MIMO system for the wide range of phase noise innovation variances. The system parameters should be set to achieve the target performance.

\section{Future Work}
 In Chapter \ref{chap:MIMO-model}, the channel gains are assumed to be known at the receiver side. The effects of the estimation errors of the channel gains on the performance of the EM-based algorithm is not investigated. The channel gains can be obtained by a conventional data-aided estimator. Additionally, the channel estimation of the block fading MIMO systems can be embedded into the  EM-based algorithms and initial estimates can be improved at every EM algorithm iteration. Therefore, this work can be extended to joint phase noise and channel estimation and data detection. In \cite{Mehrpouyan}, a data-aided LS estimator a decision-directed WLS estimator and a new decision-directed EKF is proposed. These estimators can be used in the EM-based algorithm to carry out the maximization step. Finally, statistics of the phase noise estimates can be used by taking into account in the decoding process. A receiver employing factor graphs can be implemented and to track strong phase noise processes.

\end{document}